\documentclass[fleqn,usenatbib]{mnras}

\usepackage{newtxtext,newtxmath}
\usepackage{natbib}
\usepackage{hyperref}
\usepackage{multirow}
\usepackage{longtable}
\usepackage{url}
\usepackage{mathtools}

\usepackage[dvipsnames]{xcolor}
\definecolor{myblue}{HTML}{268BD2}
\definecolor{mygreen}{HTML}{859900}
\definecolor{myred}{HTML}{DC322F}
\definecolor{mymagenta}{HTML}{D33682}

\hypersetup{
unicode=true,           % non-Latin characters in Acrobat’s bookmarks
colorlinks=true,        % false: boxed links; true: colored links
linkcolor=myred,        % color of internal links
citecolor=myblue,       % color of links to bibliography
filecolor=magenta,      % color of file links
urlcolor=mygreen        % color of external links
}

\usepackage[T1]{fontenc}
\usepackage{ae,aecompl}

\usepackage{graphicx}	% Including figure files
\usepackage{xcolor}
\usepackage{amsmath}	% Advanced maths commands
\usepackage[normalem]{ulem}

\usepackage{newtxtext,newtxmath}
\title[GRB 190114C]{Low frequency view of GRB 190114C reveals time varying shock micro-physics}

\author[K. Misra et al.]{K. Misra$^1$,\thanks{E-mail: kuntal@aries.res.in (KM)}
L. Resmi$^{2,3}$,
D. A. Kann$^{4}$,
M. Marongiu$^{5,6}$,
A. Moin$^{7}$,
S. Klose$^{8}$,
\newauthor
G. Bernardi$^{9,10,11}$,
A. de Ugarte Postigo$^{4,12}$,
V. K. Jaiswal$^2$,
S. Schulze$^{13}$,
D. A. Perley$^{14}$,
\newauthor
A. Ghosh$^{1,15}$,
Dimple$^{1,16}$,
H. Kumar$^{17,18}$,
R. Gupta$^{1,16}$,
M. J. Micha{\l}owski$^{19}$,
\newauthor
S. Mart\'in$^{20,21}$,
A. Cockeram$^{14}$,
S. V. Cherukuri$^2$,
V. Bhalerao$^{17}$,
G. E. Anderson$^{22}$,
\newauthor
S. B. Pandey$^{1}$,
G. C. Anupama$^{23}$,
C.~C.~Th\"one$^{4}$,
S. Barway$^{23}$,
M. H. Wieringa$^{24}$,
\newauthor
J.~P.~U.~Fynbo$^{25,26}$,
N. Habeeb$^{7}$
\\
$^1$ARIES, Manora Peak, Nainital 263001 India \\
$^2$Indian Institute of Space Science \& Technology, Thiruvananthapuram 695547, India\\
$^3$Anton Pannekoek Institute for Astronomy, Amsterdam 1098XH, The Netherlands \\
$^{4}$Instituto de Astrof\' isica de Andaluc\' ia (IAA-CSIC), Glorieta de la Astronom\' ia, s/n, E-18008, Granada, Spain\\
$^{5}$Department of Physics and Earth Science, University of Ferrara, via Saragat 1, I--44122, Ferrara, Italy \\
$^{6}$ICRANet, Piazzale della Repubblica 10, I--65122, Pescara, Italy \\ 
$^{7}$Department of Physics, College of Science, UAE University, Sheikh Khalifa Bin Zayed Road, Al Ain, AD UAE\\
$^{8}$Th\"uringer Landessternwarte Tautenburg, Sternwarte 5, 07778 Tautenburg, Germany \\
$^{9}$INAF - Istituto di Radioastronomia, via Gobetti 101, I-40129 Bologna, Italy\\
$^{10}$Department of Physics and Electronics, Rhodes University, PO Box 94, Grahamstown 6140, South Africa\\ 
$^{11}$ South African Radio Astronomy Observatory, Black River Park, 2 Fir Street, Observatory, Cape Town, 7925, South Africa\\
$^{12}$DARK Cosmology Centre, Niels Bohr Institute, University of Copenhagen, Lyngbyvej 2, DK-2100 Copenhagen \O, Denmark\\
$^{13}$Department of Particle Physics and Astrophysics, Weizmann Institute of Science, Rehovot 761000, Israel\\
$^{14}$ Astrophysics Research Institute, Liverpool John Moores University, 146 Brownlow Hill, Liverpool L3 5RF, United Kingdom\\
$^{15}$School of Studies in Physics and Astrophysics, Pandit Ravishankar Shukla University, Chattisgarh 492 010, India \\
$^{16}$Department of Physics, Deen Dayal Upadhyaya Gorakhpur University, Gorakhpur 273009, India\\ 
$^{17}$Indian Institute of Technology Bombay, Powai, Mumbai 400076, India\\
$^{18}$LSSTC Data Science Fellow \\
$^{19}$Astronomical Observatory Institute, Faculty of Physics, Adam Mickiewicz University, ul.~S{\l}oneczna 36, 60-286 Pozna{\'n}, Poland\\
$^{20}$European Southern Observatory, Alonso de C{\`o}rdova, 3107, Vi- tacura, Santiago 763-0355, Chile\\
$^{21}$Joint ALMA Observatory, Alonso de C{\`o}rdova, 3107, Vitacura, Santiago 763-0355, Chile\\
$^{22}$International Centre for Radio Astronomy Research, Curtin University, GPO Box U1987, Perth, WA 6845, Australia \\
$^{23}$Indian Institute of Astrophysics,  Koramangala, Bangalore 560 034, India\\
$^{24}$CSIRO Astronomy and Space Science, PO Box 76, Epping, NSW 1710, Australia\\
$^{25}$The Cosmic Dawn Center (DAWN), Niels Bohr Institute, University of Copenhagen, Lyngbyvej 2, DK-2100 Copenhagen \O, Denmark\\
$^{26}$Niels Bohr Institute, University of Copenhagen, Lyngbyvej 2, 2100 Copenhagen \O, Denmark\\
}
\date{Accepted XXX. Received YYY; in original form ZZZ}

\pubyear{2019}

\newcommand{\swift}{\textit{Swift}}
\newcommand{\fermi}{\textit{Fermi}}

\begin{document}
\label{firstpage}
\pagerange{\pageref{firstpage}--\pageref{lastpage}}
\maketitle

\begin{abstract}
We present radio and optical afterglow observations of the TeV-bright long Gamma Ray Burst (GRB) 190114C at a redshift of $z=0.425$, which was detected by the MAGIC telescope. Our observations with ALMA, ATCA, and uGMRT were obtained by our low frequency observing campaign and range from $\sim1$ to $\sim140$ days after the burst and the optical observations were done with three optical telescopes spanning up to $\sim25$ days after the burst. Long term radio/mm observations reveal the complex nature of the afterglow, which does not follow the spectral and temporal closure relations expected from the standard afterglow model. We find that the microphysical parameters of the external forward shock, representing the share of shock-created energy in the non-thermal electron population and magnetic field, are evolving with time. The inferred kinetic energy in the blast-wave depends strongly on the assumed ambient medium density profile, with a constant density medium demanding almost an order of magnitude higher energy than in the prompt emission, while a stellar wind-driven medium requires approximately the same amount energy as in prompt emission.
\end{abstract}


%%%%%%%%%%%%%%%%%%%%%%%%%%%%%%%%%%%%%%%%%%%%%%%%%%

%%%%%%%%%%%%%%%%% BODY OF PAPER %%%%%%%%%%%%%%%%%%

\section{Introduction}
Gamma-ray bursts (GRBs) dissipate between $10^{51}$ and $10^{54}$~erg (assuming isotropy) in electromagnetic radiation \citep[e.g.,][]{Amati2008a} during an ephemeral flash of $\gamma$-ray photons that can last up to thousands of s \citep[e.g.,][]{Kouveliotou1993a, Levan2014a}.
As the blastwave from the explosion sweeps up the external medium, local random magnetic fields accelerate electrons to ultra-relativistic velocities, these eventually generate a long lasting afterglow from radio to X-ray frequencies, predominantly via synchrotron radiation \citep[for a review see][]{Piran2004a, KumarZhang2014}.  Afterglow studies are an invaluable tool to answer fundamental questions on radiation processes in extreme environments. Various physical parameters such as the jet collimation angle, the state of the plasma via microphysical parameters describing magnetic field generation and electron acceleration, and the environment properties such as the density profile of the circumburst medium and dust extinction, can be constrained by modeling the multi-band evolution of the afterglow. 

The simplest afterglow model considers a power-law shape in the electron energy distribution {\it p}, a break between the optical and X-rays due to the passage of the cooling frequency and a jet with half opening angle {$\theta_j$} traversing in a constant density \citep{Schulze2011a} or wind medium \citep{1999ApJ...520L..29C}. 

However, well-sampled GRB afterglows (in the time and frequency domain) have not been found to be fully consistent with the simple afterglow model. {\it Swift} observations of the X-ray afterglows revealed plateaus,  in the lightcurves,  that can last upto $10^4$~s and whose origin continues to be poorly understood \citep{Nousek2006a, Liang2007a}. A small number of afterglows showed a rapid decline in the early optical and radio lightcurves due to the reverse shock \citep[e.g.,][]{Sari1999a, Kobayashi2003a, Laskar2013a, Martin-Carrillo2014, Gao2015a, Alexander2017b}. Others exhibited rebrightenings from X-rays to radio frequencies due to refreshed shocks \citep{Bjornsson2004a, Zhang2006a} and flares \citep{Chincarini2010a, Margutti2010a} due to on-going central-engine activity on all time scales.  But the jet geometry can also show deviations from the simplest model, a uniform top-hat jet \citep{Oates2007a, Racusin2008a, Filgas2011b} for example a two component jet \citep{Resmi2005a} and a structured jet \citep{2019ApJ...870L..15L, 2018ApJ...867...57R, 2018ApJ...856L..18M, 2018MNRAS.481.1597G}.

While most of these findings require only adjustments or additions to the standard model, a growing number of GRB afterglows start to challenge the well-established paradigm. The \fermi\ satellite recorded delayed, extended GeV emission for a number of GRBs, which may be connected to the afterglow \citep{Abdo2009a, Kumar2010a}. In the time domain, high-cadence observations of the optical and NIR afterglow of GRB 091127 pointed to a time-dependent fraction of energy stored in the magnetic field of the blastwave \citep{Filgas2011a}. But very long monitoring campaigns also revealed new challenges. \cite{DePasquale2016a} monitored the X-ray afterglow of the highly energetic GRB 130427A for $80\times10^6$~s which follows a single power-law decay. This suggested a low collimation and/or extreme properties of the circumburst medium.   \cite{2013ApJ...776..106H} have indicated two physically distinct population of GRBs - the radio bright and radio faint and suggested that this difference is due to the gamma-ray efficiency of the prompt emission between the two populations.

To understand how these examples fit into the established afterglow framework, afterglows are needed with well-sampled lightcurves from radio to X-ray frequencies. Among the $>1200$ \swift\ GRBs only $\sim20-30$ were bright enough to perform precision tests of afterglow models \citep[e.g.][]{Panaitescu2002a, Yost2003a, Bjornsson2004a, Resmi2005a, Chandra2008a, Laskar2013a, Sanchez2017a, Alexander2017b}. With the dawn of the Atacama Large Millimeter/submillitmeter Array (ALMA; \citealt{Wootten2009a}), upgraded Giant Metre-wave Radio Telescope (uGMRT; \citealt{Swarup1991a}), the Karl G. Jansky Very Large Array (VLA),
and the NOrthern Extended Millimeter Array (NOEMA), it is finally feasible to not only monitor bright GRBs over a substantially longer period of time, but also less luminous and more distant GRBs. In addition, analytical and sophisticated numerical models, e.g. \citet{Johaneesson2006a}, \citet{vanderHorst2007a}, \citet{vanEerten2012}, and \citet{Laskar2013a, Laskar2015a}, were also revised to account for the observed afterglow diversity.

On 14 January 2019, the MAGIC air Cherenkov telescope recorded for the first time very high-energy (VHE) photons from a GRB, GRB 190114C. This provides an opportunity to study not only leptonic but also hadronic processes in GRBs and their afterglows. In this paper, we present the results of our observing campaign of the afterglow of GRB 190114C with ATCA, ALMA and GMRT at radio frequencies and with the 0.7m GROWTH-India telescope (GIT), the 1.3m Devasthal Fast Optical Telescope (DFOT), and the 2.0m Himalayan Chandra Telescope (HCT)
in the optical bands. A brief description of the burst properties is given in Sect. \ref{grb190114c}. The data acquisition and analysis procedures are described in Sect. \ref{data}.  The multi-band afterglow lightcurves and the description of the afterglow in the context of other GRB afterglows are discussed in Sect. \ref{section:3}. We discuss the interstellar scintillation in the radio bands in Sect. \ref{par:scint}.  A detailed multi-band modelling of the afterglow lightcurves reveals the evolution of microphysical parameters with time, as presented in Sect. {\ref{section:4}}.  The conclusions of this work are given in Sect. \ref{conclusions}.  The time since burst (T-T$_0$) is taken to be the \swift\ trigger time. We adopt the convention of $F_{\nu}(t) \propto t^{\alpha}\nu^{\beta}$ throughout the description given in this work.

Throughout the paper, we report all uncertainties at $1\sigma$ confidence and the brightness in the UV/optical/NIR in the AB magnitude system. We use a $\Lambda$CDM cosmology with $H_0=67.3~{\rm km\,s}^{-1}\,{\rm Mpc}^{-1}$, $\Omega_\Lambda=0.685$ and $\Omega_{\rm m}=0.315$ \citep{Planck2014a}. 

%\newpage\clearpage

\section{The MAGIC burst GRB 190114C}
\label{grb190114c}
GRB 190114C was first detected by the Burst Alert Telescope \citep[BAT,][]{2005SSRv..120..143B} onboard the \textit{Neil Gehrels Swift Observatory} satellite \citep[hereafter \textit{Swift},][]{2004ApJ...611.1005G} on January 14, 2019 at 20:57:03.19 UT  with a $T_{\rm 90}$ duration of 25 sec (\citealt{2019GCN.23707....1H}, \citealt{2019GCN.23688....1G}). The GRB was also detected by other high energy missions such as the SPI-ACS detector onboard \textit{INTEGRAL} which recorded prolonged emission up to $\sim5000$ s \citep{2019GCN.23714....1M},  {\it Insight-HXMT} \citep{2019GCN.23716....1X}, Konus-\textit{Wind}, which recorded emission in the 30 keV to 20 MeV energy band \citep{2019GCN.23737....1F}, as well as the GBM and LAT instruments onboard the \fermi\ satellite, with the highest-energy photon detected at 22.9 GeV 15 s after the GBM trigger.

A historically rapid follow-up observation, $\sim50$ s after the BAT trigger, of GRB 190114C was performed by the twin Major Atmospheric Gamma Imaging Cherenkov (MAGIC) telescopes \citep{2019GCN.23701....1M,2019GCN.23688....1G}. The MAGIC real-time analysis detected very high energy emission $>300$ GeV with a significance of more than $20\sigma$ in the first 20 minutes of observations. The higher detection threshold comes due to the large zenith angle of the observation ($>60$ deg) and the presence of a partial Moon. However, after an initial flash of very high energy gamma-ray photons, the VHE emission quickly faded, as expected for a GRB and corroborating the connection between the VHE flash with the GRB.

Furthermore, the \swift\ X-ray Telescope \citep[XRT,][]{2005SSRv..120..165B} started observing the field 64 s after the BAT trigger and located an uncatalogued X-ray source \citep{2019GCN.23688....1G}.  The UV/optical afterglow was also detected by the \swift\ UV/Optical Telescope \citep[UVOT,][]{2005SSRv..120...95R} 73 s after the BAT trigger.

A series of optical observations were obtained with several telescopes, such as the Master-SAAO robotic telescope \citep{lipunov1}, the 2.5 m NOT \citep{2019GCN.23695....1S}, the 0.5 m OASDG \citep{2019GCN.23699....1I}, as well as the 2. m MPG/ESO telescope with GROND \citep{Greiner2008PASP} which detected the afterglow in multiple filters \citep{2019GCN.23702....1B}. A redshift of $z=0.425$ \citep{2019GCN.23695....1S} was measured from the strong absorption lines seen in the spectrum taken with the ALFOSC instrument on the 2.5m NOT.  This was further refined ($z=0.4245\pm0.0005$) and confirmed by GTC \citep{2019GCN.23708....1C} and VLT X-shooter \citep{2019GCN.23710....1K}  spectroscopic observations. The measured fluence by the \fermi\ GBM is $3.99 \times 10^{-4} \pm 8 \times 10^{-7}$ erg/cm$^{2}$ in the 10-1000 keV energy range, hence the total isotropic energy and isotropic luminosity of the burst are $E_{\rm iso} \sim 3 \times 10^{53}$ erg and $L_{\rm iso} \sim 1 \times 10^{53}$ erg/s respectively \citep{2019GCN.23707....1H}. The values for this burst are in agreement with the $E_{\rm peak}-E_{\rm iso}$ \citep{2002A&A...390...81A} and $E_{\rm peak}-L_{\rm iso}$ \citep{2004ApJ...609..935Y} correlations.

\section{Data acquisition and analysis}
\label{data}
\subsection{Optical observations}
We undertook photometric observations of GRB 190114C with the 0.7-m GROWTH India telescope (GIT) and the 2.0m Himalayan Chandra Telescope (HCT), located at the Indian Astronomical Observatory (IAO), Hanle, India, as well as the 1.3m Devasthal Fast Optical Telescope (DFOT), located in Devasthal, ARIES, Nainital, India. The first observations with GIT were performed about 16.36~hrs after the initial alert. A faint afterglow was detected in $g^\prime$, $r^\prime$, and $i^\prime$ filters \citep{ksw+19}. We monitored the afterglow upto 25.71 days after the trigger. The GIT is equipped with a wide-field camera with large ($\sim0\farcs7$) pixels, and typical stars in our images in this data set had a full-width-at-half-maximum (FWHM) of 4\arcsec.  Image processing including bias subtraction, flat-fielding and cosmic-ray removal was done using standard tasks in {\sc IRAF} \citep{1993ASPC...52..173T}. Source Extractor (SExtractor, \citealt{bertin11}) was used to extract the sources in GIT images. The zero points were calculated using PanSTARRS reference stars in the GRB field.  In DFOT and HCT images, psf photometry was performed to estimate the instrumental magnitudes of the afterglow which were calibrated against the same set of stars as used in GIT using USNO catalog.  The final photometry is listed in Table~\ref{tab:gitphot}.

\begin{table}
\caption{Optical photometry of the afterglow of GRB~190114C using GIT, DFOT and HCT. The magnitudes are not corrected for Galactic extinction. The GIT magnitudes are in the AB and the DFOT and HCT magnitudes are in the Vega system.}
\begin{center}
\begin{tabular}{c c c c c} 
\hline
T-T$_0$ & Filter & Magnitude & Telescope & Image FWHM \\  
(days) & &  & & (arcsec) \\
\hline
0.677 & $g^\prime$ & $20.61\pm0.32$ & GIT &  3.63 \\
0.697 & $r^\prime$ & $19.46\pm0.07$ & GIT & 3.23 \\
0.707 & $i^\prime$ & $19.48\pm0.16$ & GIT & 3.77 \\
0.724 & R         &  $19.39\pm0.06$ & DFOT &  3.18\\     
0.729 & R          &  $19.46\pm0.07$ & DFOT &  3.25 \\
0.748 & R          &  $19.70\pm0.10$ & HCT &  2.18  \\                   
0.803 &  R          &  $19.81\pm0.09$ & HCT & 2.10 \\
1.827 & $r^\prime$ & $19.96\pm0.20$ & GIT & 3.27 \\
1.839 & $r^\prime$ & $20.04\pm0.26$ & GIT & 3.44 \\
2.715 & $r^\prime$ & $20.71\pm0.27$ & GIT & 3.57 \\
2.782 & $i^\prime$ & $19.78\pm0.24$ & GIT & 3.21 \\
3.819 & $i^\prime$ & $19.53\pm0.45$ & GIT & 4.18 \\
4.813 & $r^\prime$ & $20.23\pm0.37$ & GIT & 4.51 \\
9.753 & R         & $21.38\pm0.16$ & HCT & 2.73 \\
9.757 & $r^\prime$ & $21.53\pm0.32$ & GIT & 3.98 \\
9.791 & $i^\prime$ & $20.35\pm0.39$ & GIT & 4.04 \\
10.788 & $g^\prime$ & $21.96\pm0.40$ & GIT& 4.08 \\ 
10.801 & $r^\prime$ & $21.35\pm0.41$ & GIT& 4.01 \\
10.813 & $i^\prime$ & $20.05\pm0.37$ & GIT& 4.05 \\
11.770 & $r^\prime$ & $21.44\pm0.45$ & GIT& 4.35 \\
11.782 & $i^\prime$ & $20.32\pm0.36$ & GIT& 4.07 \\
12.757 & $r^\prime$ & $21.65\pm0.35$ & GIT& 4.85 \\
12.768 & $r^\prime$ & $21.74\pm0.41$ & GIT& 4.95 \\
12.800 & $i^\prime$ & $20.11\pm0.22$ & GIT& 3.81 \\
13.703 & $g^\prime$ & $21.70\pm0.40$ & GIT& 4.04 \\
13.726 & $r^\prime$ & $21.56\pm0.36$ & GIT& 4.03 \\
13.736 & $r^\prime$ & $21.44\pm0.28$ & GIT& 4.03 \\
13.747 & $i^\prime$ & $20.29\pm0.24$ & GIT& 3.97 \\
15.692 & $r^\prime$ & $21.35\pm0.31$ & GIT& 4.12 \\
15.728 & $i^\prime$ & $20.59\pm0.35$ & GIT& 4.01 \\
18.736 & $i^\prime$ & $20.76\pm0.33$ & GIT& 3.97 \\
19.772 & $i^\prime$ & $20.57\pm0.24$ & GIT& 3.72 \\
24.709 & $r^\prime$ & $21.26\pm0.30$ & GIT& 4.09 \\
24.724 & $i^\prime$ & $20.42\pm0.30$ & GIT& 4.09 \\
25.697 & $r^\prime$ & $21.25\pm0.31$ & GIT& 4.17 \\
25.714 & $i^\prime$ & $20.70\pm0.40$ & GIT& 4.18 \\ 
\hline
\end{tabular}
\end{center}
\label{tab:gitphot}
\end{table}
 
\subsection{ATCA}
Radio observations of the GRB 190114C \swift\ XRT position \citep{Osborne2019GCN23704} were carried out with the Australia Telescope Compact Array (ATCA), operated by CSIRO Astronomy and Space Science under a joint collaboration team project (project code CX424, \citealt{2019GCN.23745....1S}).  Data were obtained using the CABB continuum mode \citep{Wilson2011a} with the 4 cm (band centres: 5.5 and 9 GHz), 15 mm (band centres: 17 and 19 GHz) and 7 mm receivers (band centres 43 and 45 GHz), which provided two simultaneous bands, each with 2 GHz bandwidth.  Data reduction was done using the {\sc Miriad} \citep{Sault1995a} and {\sc Common Astronomy Software Applications} (CASA, \citealt{McMullin2007a}) software packages and standard interferometry techniques were applied. Time-dependent gain calibration of the visibility data was performed using the ATCA calibrator sources 0237-233 (RA = 02:40:08.17, Dec. = -23:09:15.7) or 0402-362 (RA = 04:03:53.750, Dec. = -36:05:01.91), and absolute flux-density calibration was carried out on primary ATCA flux calibrator PKS B1934-638 \citep{Partridge2016a}. Visibilities were inverted using standard tasks to produce GRB 190114C field images. The final flux-density values were estimated by employing model-fitting in both image and visibility planes to check for consistency. Table \ref{atca} shows the epochs of ATCA observations, frequency bands, the observed flux densities along with the errors and the telescope configuration during the observations. The quoted errors are $1\sigma$, which include the RMS and Gaussian $1\sigma$ errors.

%ATCA observing log
\begin{table}
\caption{ATCA observing log of the radio afterglow of GRB 190114C.}
\centering
\smallskip
\begin{tabular}{c c c c}
\hline
T-T$_0$      & Frequency & Flux density & Configuration \\ 
(days)       & (GHz)     &  (mJy)       &  \\
\hline  
3.291  &	45	& $1.311\pm0.101$ & H75 \\
9.334  &	45	& $0.547\pm0.127$ & H75 \\
10.331 &	45	& $0.516\pm0.143$ & H75 \\
12.321 &	45	& $<0.113$ & H75 \\
20.377 &	45	& $<0.094$ & H75 \\
\hline
3.291  &	43	& $1.491\pm0.085$ & H75 \\
9.334  & 	43	& $0.583\pm0.084$ & H75\\
10.331 &	43	& $0.534\pm0.110$ & H75\\
12.321 &	43	& $0.403\pm0.104$ & H75\\
20.377 &	43	& $0.362\pm0.088$ & H75\\
\hline
3.322  &	19	& $2.810\pm0.059$ & H75\\
5.332  &	19	& $2.440\pm0.200$ & H75\\
9.366  &	19	& $2.730\pm0.054$ & H75\\
10.362 &    19	& $2.000\pm0.070$ & H75\\
16.328 &	19	& $1.180\pm0.031$ & H75\\ 
35.430 &	19	& $0.732\pm0.028$ & H75\\
\hline
1.455  & 	18 &	$2.000\pm0.800$ & 1.5D \\
3.322  &	18 &	$2.530\pm0.280$ & H75\\
5.332  &	18 &	$2.060\pm0.260$ & H75\\
9.366  &	18 &	$1.820\pm0.078$ & H75\\
10.362 &	18 &    $1.820\pm0.120$ & H75\\
16.328 &	18 &	$0.770\pm0.050$ & H75\\
35.430 &	18 &	$0.520\pm0.090$ & H75\\
52.338 &	18 &    $<0.48$ & \\
\hline
3.322  &	17	& $3.180\pm0.037$ & H75\\
5.332  &	17	& $2.995\pm0.068$ & H75\\
9.366  &	17	& $2.081\pm0.033$ & H75\\
10.362 &	17	& $2.130\pm0.030$ & H75\\
12.352 &	17  & $1.050\pm0.150$ & H75\\
16.328 &    17	& $1.142\pm0.010$ & H75\\
24.422 &	17  & $0.560\pm0.080$ & H75\\
35.430 &	17	& $0.614\pm0.032$ & H75\\
\hline
1.424  & 9.0	&	$1.820\pm0.040$ & 1.5D \\
3.489  & 9.0	&	$2.080\pm0.040$ & H75\\
5.301  & 9.0	&	$2.230\pm0.045$ & H75\\
9.334  & 9.0	&	$1.580\pm0.017$ & H75\\
10.331 & 9.0	&	$1.500\pm0.021$ & H75\\
16.297 & 9.0	&	$0.927\pm0.019$ & H75\\
20.377 & 9.0	&	$0.802\pm0.016$ & H75\\
24.262 & 9.0	&	$0.560\pm0.050$ & H75\\
35.315 & 9.0	&	$0.420\pm0.020$ & H75\\
52.307 & 9.0	&	$0.240\pm0.020$ & H214 \\
72.515 & 9.0	&	$0.150\pm0.010$ & 6A \\
119.282 & 9.0	&	$0.108\pm0.015$ & 1.5B \\
138.217	& 9.0	&	$0.094\pm0.014$ & 6A \\
\hline
1.424   & 5.5 &	$1.930\pm0.030$ & 1.5D \\	
3.489	& 5.5 &	$1.140\pm0.030$ & H75\\	
5.301	& 5.5 &	$1.770\pm0.037$ & H75\\	
9.334	& 5.5 &	$2.210\pm0.032$ & H75\\	
10.331	& 5.5 &	$1.200\pm0.031$ & H75\\	
16.297	& 5.5 &	$0.735\pm0.023$ & H75\\	
20.377	& 5.5 &	$0.720\pm0.018$ & H75\\	
24.262	& 5.5 &	$0.670\pm0.040$ & H75\\	
35.315	& 5.5 &	$0.480\pm0.030$ & H75\\	
52.307	& 5.5 &	$0.340\pm0.020$ & H214 \\	
72.515	& 5.5 &	$0.240\pm0.020$ & 6A \\	
119.282	& 5.5 &	$0.140\pm0.015$ & 1.5B \\	
138.217	& 5.5 &	$0.126\pm0.013$ & 6A\\
\hline
\end{tabular}
\label{atca}
\end{table} 

\subsection{ALMA}
The afterglow of GRB 190114C was observed with the Atacama Large Millimetre/Submillimetre Array (ALMA) in Bands 3 and 6. These observations were performed between 15 January and 1 March 2019 (1.1 and 45.5 days after the burst). The angular resolution of the observations ranged between 2\farcs58 and 3\farcs67 in Band 3 and was 1\farcs25 in Band 6.

Band 6 observations were performed within the context of DDT programme ADS/JAO.ALMA\#2018.A.00020.T (P.I.: de Ugarte Postigo). Five individual executions were performed in three independent epochs ranging between January 17 and 18, 2019. The configuration used $47-48$ antennas with baselines ranging from 15~m to 313~m ($12-253~k\lambda$ at the observed frequency). Each observation consisted of 43 min integration time on source with average weather conditions of precipitable water vapour (pwv) $\sim3-4$~mm.  The receivers were tuned to a central frequency of 235.0487 GHz, so 
that the upper side band spectral windows would cover the CO(3-2) transition at the redshift of the GRB. The spectroscopic analysis of these data was presented by \citet{deUgartePostigo2019}, whereas in this paper we make use of the continuum measurements. The spatial resolution of the spectral data cube obtained by the pipeline products that combined all five executions was $1\farcs16\times0\farcs867$ (Position Angle $-87.9^\circ$).

The ALMA Band 3 observations were performed within the ToO programme ADS/JAO.ALMA\#2018.1.01410.T (P.I.: Perley) on January 15, 19, and 25, and on March 1 following the annual February shutdown.  Integration times were 8.6 minutes on-source per visit.  Weather conditions were relatively poor, with pwv~$3-4$~mm (accompanying Band 7 observations were requested, but could not be executed under the available conditions).

All data were calibrated within CASA version 5.4.0 using the pipeline calibration. Photometric measurements were also performed within CASA. The flux calibration was performed using J0423-0120 (for the first Band 3 epoch and the last Band 6 epoch) and J0522-3627 (for the remaining epochs).  The log of ALMA observations and flux density measurements along with the errors are given in Table \ref{alma}.

%ALMA observing log
\begin{table}
\caption{ALMA observing log of the afterglow of GRB 190114C. For each epoch we provide photometric measurements performed in the four side bands, as well as the combined photometry of all the bands, indicated with a $*$. \label{tab:log}}
\centering
\smallskip
\begin{tabular}{c c c}
\hline
 T-T$_0$     & Frequency & Flux density     \\
(days)       & (GHz)     & (mJy)         \\
\hline  
%17-Jan-2019/02:12:43.8   to   17-Jan-2019/03:14:43.8 (UTC) --> Xc8a [d]
2.240       & 227.399 & $1.459\pm0.043$\\
2.240       & 229.505 & $1.443\pm0.045$\\
2.240       & 240.581 & $1.559\pm0.048$\\
2.240       & 242.698 & $1.480\pm0.057$\\
2.240$^{*}$ & 235.048 & $1.480\pm0.026$\\ % All Combined
\hline
%17-Jan-2019/03:39:50.4   to   17-Jan-2019/04:41:49.5 (UTC) --> X106d [a] [This epoch looks a bit worse]
2.301       & 227.399 & $1.180\pm0.056$\\
2.301       & 229.505 & $1.261\pm0.061$\\
2.301       & 240.581 & $1.132\pm0.061$\\
2.301       & 242.697 & $1.475\pm0.071$\\ 
2.301$^{*}$ & 235.048 & $1.267\pm0.034$\\ % All Combined
\hline
%18-Jan-2019/01:27:52.7   to   18-Jan-2019/02:30:03.0 (UTC) --> X6816 [b]
3.209       & 227.399 & $1.468\pm0.092$\\
3.209       & 229.505 & $1.459\pm0.099$\\
3.209       & 240.581 & $1.382\pm0.044$\\
3.209       & 242.698 & $1.510\pm0.048$\\
3.209$^{*}$ & 235.048 & $1.462\pm0.020$\\ % All Combined
\hline
%118-Jan-2019/02:31:53.5   to   18-Jan-2019/03:33:54.3 (UTC) --> X6dcc [c]
3.254      & 227.399 & $1.340\pm0.045$\\
3.254      & 229.505 & $1.338\pm0.043$\\
3.254      & 240.581 & $1.321\pm0.093$\\
3.254      & 242.698 & $1.360\pm0.048$\\
3.254$^{*}$ & 235.048 & $1.322\pm0.025$\\ % All combined
\hline
%18-Jan-2019/22:39:28.5   to   18-Jan-2019/23:44:12.6 (UTC) --> Xebe1 [e]
4.093       & 227.399 & $1.300\pm0.041$\\
4.093       & 229.505 & $1.442\pm0.041$\\
4.093       & 240.581 & $1.386\pm0.057$\\
4.093       & 242.698 & $1.495\pm0.059$\\
4.093$^{*}$ & 235.048 & $1.393\pm0.026$\\ % All Combined
\hline
%2019-01-15T22:52:11.904000								
1.080       & 90.5    & $2.668\pm0.051$\\			
1.080       & 92.5	  & $2.758\pm0.053$\\				
1.080       & 102.5	  & $2.533\pm0.052$\\				
1.080       & 104.5	  & $2.524\pm0.056$\\			
1.080$^{*}$ & 97.5 	  & $2.618\pm0.027$\\ %Continuum
\hline
%2019-01-19T00:03:37.680000								
4.130       & 90.5	 & $1.711\pm0.052$\\				
4.130       & 92.5   & $1.764\pm0.049$\\				
4.130       & 102.5  & $1.618\pm0.050$\\				
4.130       & 104.5  & $1.557\pm0.052$\\			
4.130$^{*}$ & 97.5   & $1.682\pm0.029$\\ %Continuum
\hline
%2019-01-25T02:43:01.488000								
10.240      & 90.5   & $1.189\pm0.054$\\				
10.240      & 92.5   & $0.948\pm0.062$\\			
10.240      & 102.5  & $0.984\pm0.055$\\				
10.240      & 104.5  & $0.911\pm0.060$\\				
10.240$^{*}$& 97.5   & $1.005\pm0.032$\\ %Continuum
\hline
45.5$^{*}$  & 97.5   & <0.14 \\ 
\hline
\end{tabular}
\label{alma}
\end{table}
 
 \subsection{GMRT}
 We observed GRB 190114C in band-4 ($550-850$~MHz) and band-5 ($1050-1450$~MHz) of the upgraded Giant Metrewave Radio Telescope (uGMRT)  between  17 January to 25 March, 2019 ($\sim 2.8$ to 68.6 days since burst) under the approved ToO program 35\_018 (P.I.: Kuntal Misra). Either 3C147 or 3C148 was used as flux calibrator and 0423-013 was used as phase calibrator.
 
 We used a customised pipeline developed in {\sc CASA} by Ishwar-Chandra et al. (2021, in preparation) for analysing the data. For both band-4 and band-5, about $26^{\prime}-28^{\prime}$ large regions centred on the GRB coordinates were imaged for the analysis, with a cell-size of 1\farcs24 and 0\farcs5 respectively. To measure the flux at the GRB position we used the task \textsl{JMFIT} in the Astronomical Image Processing System (AIPS). We ran the fits on a small region around the radio transient position as measured by the VLA \citep{2019GCN.23726....1A},  using a two-component model consisting of an elliptical Gaussian and a flat baseline function. On all images except the last two epochs of band-4, the fitting procedure resulted in a confident detection of an unresolved point source at the GRB position. See Table \ref{gmrt} for the observation log. Errors quoted are obtained from the {\textsl{JMFIT}} fitting routine. The upper limits for the last two epochs of band-4 observations correspond to 3 times the mean flux measured in an empty region of the map.  The synthesised beam is typically $(5-8)^{\prime\prime} \times 3^{\prime\prime}$ for the maps. The value presented for the band-5 observation on 17 January 2019 is an improvement of the measurement reported in \cite{2019GCN.23762....1C} and \cite{MagicMWLpaper}, which was from a preliminary analysis. Self-calibration of the data in our refined analysis improved the quality of the image and the confidence of the detection. The measurements are not corrected for the host-galaxy which was detected in the pre-explosion images obtained with MeerKAT data \citep{Tremou2019GCN23760}.

%GMRT observing log
\begin{table}
\caption{uGMRT observing log of the afterglow of GRB 190114C. Measurements are not corrected for the flux of the host galaxy detected in MeerKAT pre-explosion images.}
\centering
\smallskip
\begin{tabular}{c c c}
\hline
T-T$_0$      & Frequency & Flux density\\ 
(days)       & (GHz)     &  (mJy)         \\
\hline  
2.815  &	1.26	& $0.144\pm0.017$ \\
9.710  &	1.26	& $0.303\pm0.012$ \\
49.690 &	1.26	& $0.153\pm0.015$ \\
68.669 &	1.26	& $0.162\pm0.012$ \\
\hline
21.752 &	0.65 &	$0.104\pm0.030$ \\
48.544 &    0.65 &	$<0.12$ \\
66.565 &	0.65 &	$<0.18$ \\
\hline
\end{tabular}
\label{gmrt}
\end{table}

\section{Multi-band lightcurves}
\label{section:3}
\begin{figure*}
\centering
\includegraphics[width=1\textwidth]{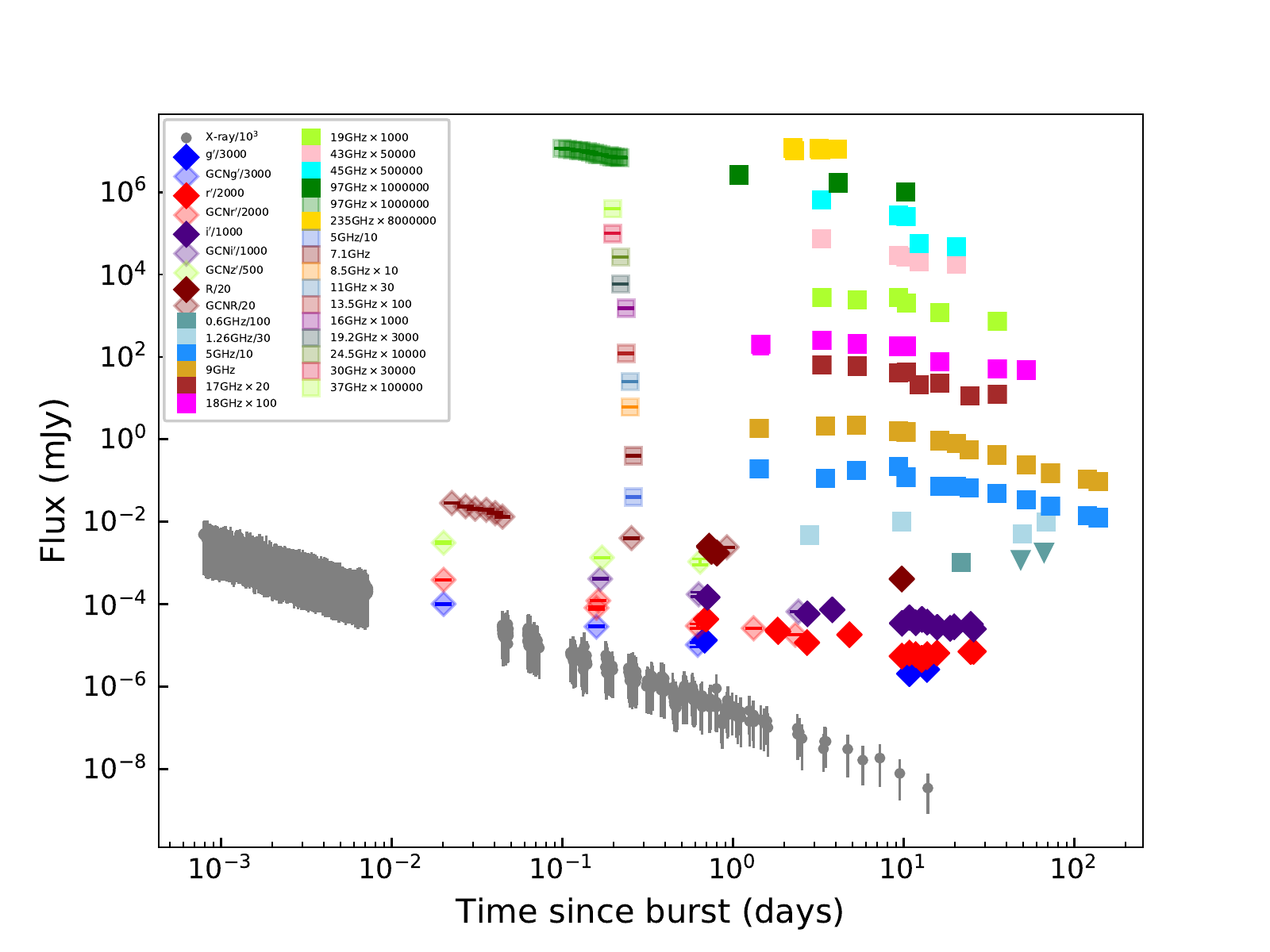}
\caption{Multi-band lightcurves of the afterglow of GRB 190114C from X-ray to the radio/mm bands. The solid symbols represent our data and the light shaded symbols represent the data from the literature.} 
\label{multibandlc}
\end{figure*}

\begin{figure}
\centering
\includegraphics[width=1\columnwidth]{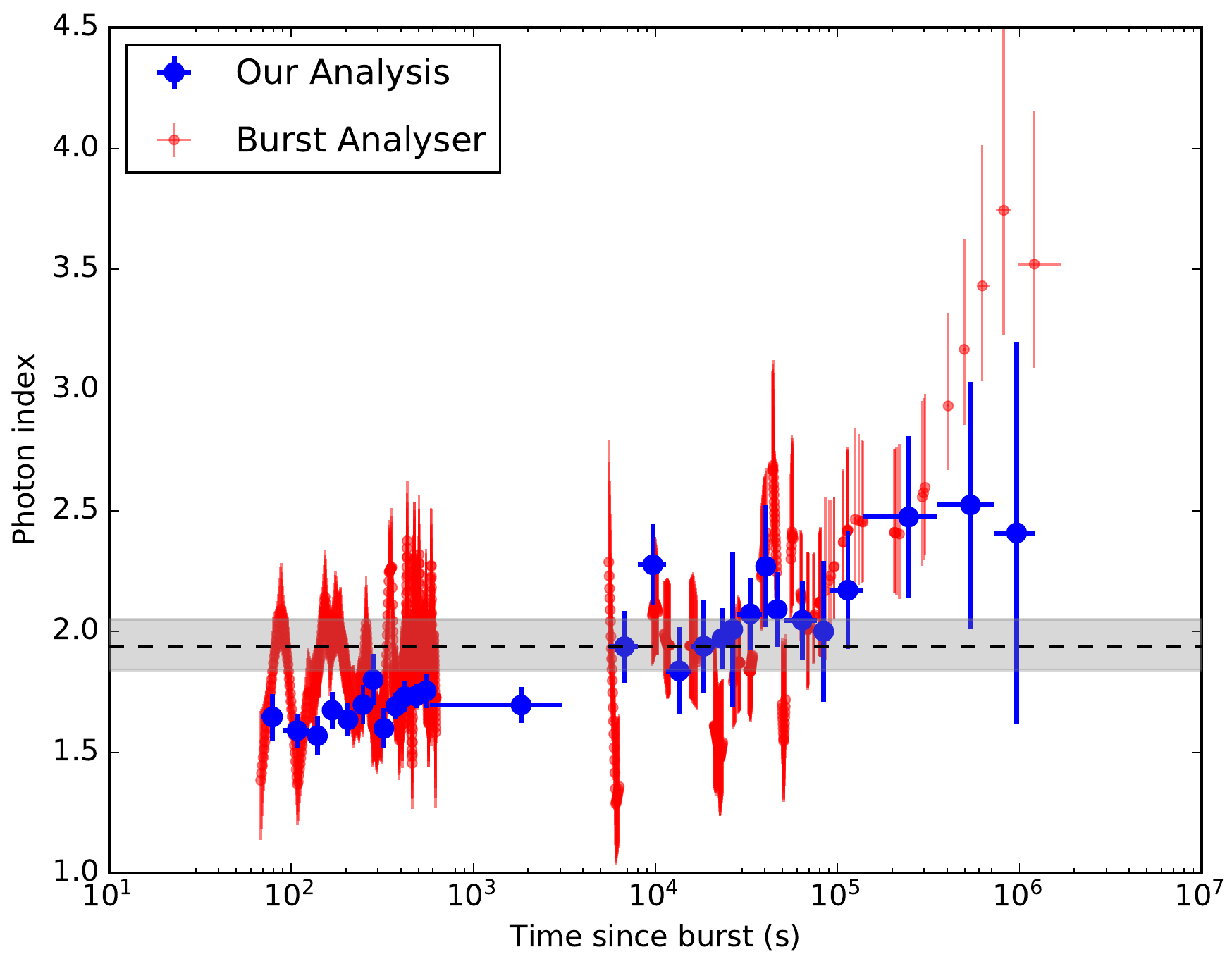}
\caption{Comparison of the X-ray photon index values from burst analyser (red) and our estimates (blue) along with the photon index value (grey horizontal bar) reported in the XRT repository.} 
\label{xray_specevolution}
\end{figure}

In Fig. \ref{multibandlc} we show the multi-band evolution of the afterglow of GRB 190114C constructed using our ATCA, ALMA, GMRT, GIT, DFOT, and HCT data and supplemented with the X-ray lightcurve obtained from the \swift\ XRT archive\footnote{\href{http://www.swift.ac.uk/xrt_curves/}{http://www.swift.ac.uk/xrt\_curves/}} \citep{2007A&A...469..379E, 2009MNRAS.397.1177E} and using available optical and radio data in the literature 
\citep{Mazaeva2019GCN23741, 
2019GCN.23734....1K, 
2019GCN.23729....1D, 
Kim2019GCN23732, 
2019GCN.23727....1M, 
2019GCN.23695....1S, 
2019GCN.23702....1B, 
2019GCN.23798....1S, 
2019GCN.23787....1M, 
Watson2019GCN23751,
2019GCN.23749....1W, 
2019GCN.23748....1R, 
2019GCN.23746....1M,
2019GCN.23742....1K,
2019GCN.23699....1I, 
Laskar2019}. 
We adopt the X-ray lightcurve from the \swift\ XRT archive considering that there is no photon index evolution.  However, The \swift~ Burst Analyser page 
\footnote{\href{https://www.swift.ac.uk/burst_analyser/}{https://www.swift.ac.uk/burst\_analyser/}}  \citep{2010A&A...519A.102E} for this burst shows that the X-ray photon index changes over time between 10$^5$ to 10$^6$ sec. In order to address this issue, we performed our own spectral fits to estimate the photon index. We binned the count rate light curve using Bayesian block binning \citep{1998ApJ...504..405S} and generated the spectra in these time bins.  The spectra are fitted with an absorbed simple power law model ({\it{phabs}}) using two absorption components - one for Galactic and one for host.  The $N_H$ values for our galaxy and host are fixed at 7.54$\times$10$^19$ cm$^{\rm -2}$ and 8.0$\times$10$^{22}$ cm$^{\rm -2}$ (taken from XRT spectrum repository) respectively and $z$ of the absorber is fixed at $0.42$. We compare our estimated photon index with that of the burst analyser.  Our results are consistent with the photon index of 1.94 (+0.11,-0.10) within error bars given in the XRT repository and no spectral evolution is evident. A comparison of the photon index values from burst analyser and our estimates along with the XRT repository photon index is shown in Fig. \ref{xray_specevolution}. Based on these results we adopt the XRT repository light curve for the rest of the analysis presented in this paper.

The multi-band evolution of the GRB 190114C afterglow is complex as seen from Fig. \ref{multibandlc}. The temporal evolution of the X-ray lightcurve is consistent with a single power-law following a decay index of $\alpha_\textnormal{X}=-1.344\pm0.003$ from 68 sec to $\sim 10$ days, and shows a hint of a steeper decline thereafter.

\begin{figure*}
\centering
\includegraphics[width=1\textwidth]{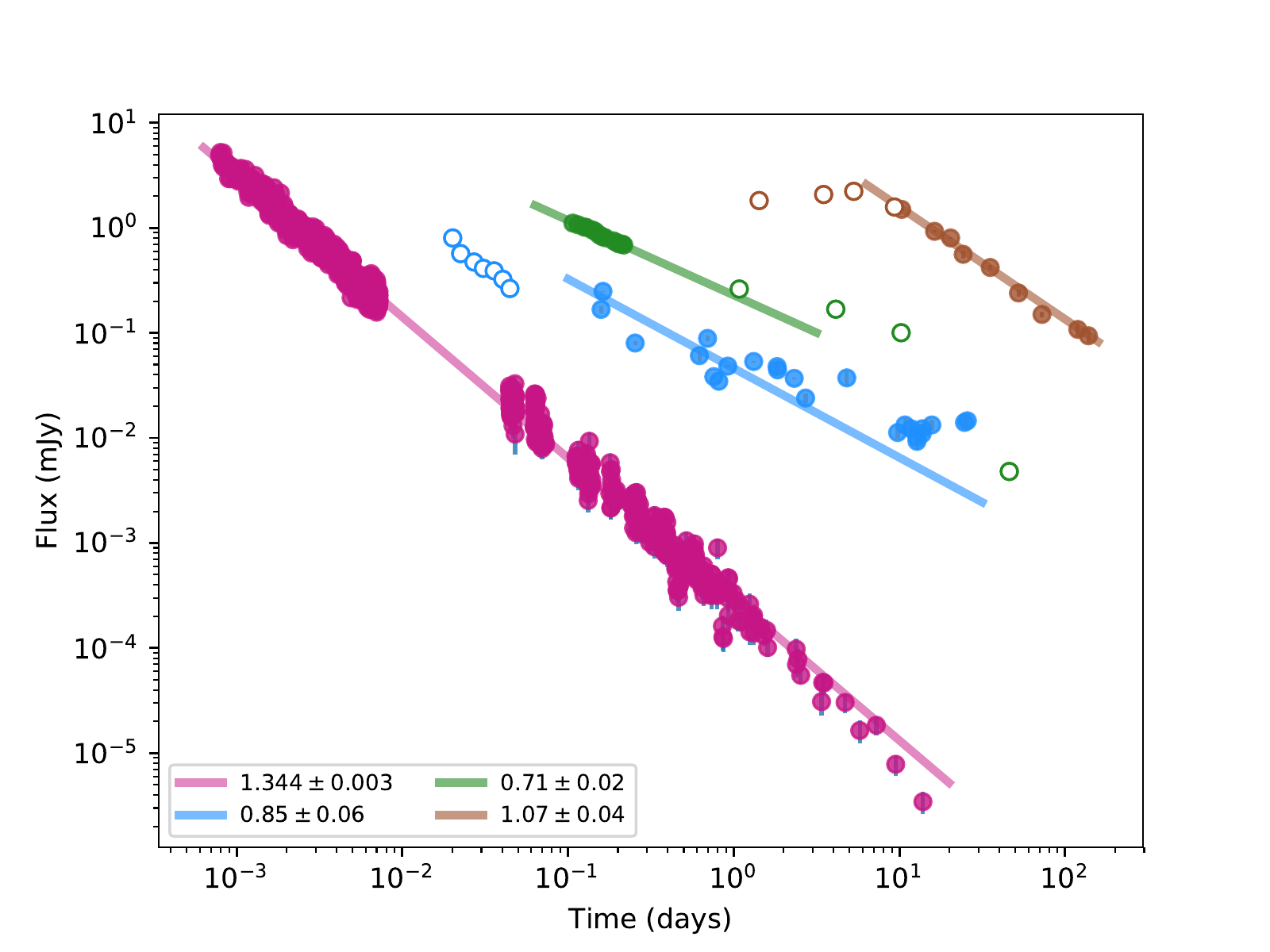}
\caption{Empirical fits to a selected set of multiband lightcurve data, used to aid the physical modelling (see text). X-ray (violet), $R$ band (blue), $97$~GHz (green), and $9$~GHz (brown). For $9$~GHz, $97$~GHz and the $R$ band, data excluded from the fit are shown with open symbols.}
\label{newfig1}
\end{figure*}

ATCA 9 and 5.5~GHz data offer a temporal coverage of two orders of magnitude. The late-time temporal slope of $9$~GHz ($t>10$~days) is $-1.07\pm0.04$ and of $5.5$~GHz ($t>25$~days) is $-1.00\pm0.03$. Millimeter data presented in this paper along with that of \cite{Laskar2019} give a wide temporal coverage at $97$~GHz. For $t < 0.3$~days the $97$~GHz lightcurve decays as $t^{-(0.71 \pm 0.02)}$, the temporal coverage is sparse afterwards, however our last observation yielding a $3\sigma$ upper limit of $0.14$~mJy indicates a steeper decay. In Fig. \ref{newfig1}, we present the power-law fits in the X-ray, $R$, $97$~GHz, and $9$~GHz bands mentioned above.

To construct a broadband multi-colour optical/NIR lightcurve, we take data from the following sources: this paper, \cite{MagicMWLpaper}; \cite{deUgartePostigo2019}; \cite{Jordana-Mitjans2020}; Melandri et al., in prep.;  and GCNs \citep{Kim2019GCN23732,Mazaeva2019GCN23741,Im2019GCN23757,Watson2019GCN23751,Bikmaev2019GCN23766}. We correct for Galactic extinction along the line-of-sight (E(B-V) = 0.0107$\pm$0.0004 mag;   \citealt{Schlafly2011ApJ}), remove outliers and fit the data set, spanning from the $uvw2$ to the $K$ band, with a set of smoothly broken power-laws. Hereby we assume achromaticity\footnote{Note that \protect\cite{Jordana-Mitjans2020} find evidence at early times for colour evolution. This increases the $\chi^2$ of our fit but does not significantly influence the SED.} and share the fit parameters early steep slope $\alpha_s$, pre-break slope $\alpha_1$, post-break slope $\alpha_2$, break times $t_{b,s}$, $t_b$ and break smoothness $n_s$, $n$ among all bands, whereas the normalisations and host-galaxy magnitudes are individual parameters for each band. We exclude any data beyond seven days (except for late host observations at $>50$ days), as they may be influenced by a rising supernova component.

We find that the earliest observations are far brighter than a back-extrapolation of the data beyond 0.1 days, likely due to a steeply decaying reverse shock component \protect\citep{Jordana-Mitjans2020}. Fitting the data up to 0.8 days yields ($\chi^2{\textnormal{/d.o.f.}} = 1.86$) $\alpha_s=-2.076\pm0.023$, $\alpha_1=-0.544\pm0.011$, and $t_{b,s}=0.00508\pm0.0003$ days ($t_{b,s}=439\pm26$ s); hereby $n=-0.5$ was fixed, a soft steep-to-shallow transition between slopes.

Between 0.06 and 7 days, the multi-colour lightcurve is well-described (some remaining scatter leads to $\chi^2{\textnormal{/d.o.f.}} = 3.35$) by a broken power-law with $\alpha_1=-0.530\pm0.017$, $\alpha_2=-1.067\pm0.011$, and $t_b=0.576\pm0.028$ days; hereby $n=10$ was fixed. The normalisation of each band, formally the magnitude at break time for $n=\infty$ \citep{Zeh2006ApJ}, then represents the UV/optical/NIR Spectral Energy Distribution (SED), based not just on a small number of data points, but on all data involved in the fit. As only the first fit covers all bands, we use the values derived from it. The direct values are measured at break time $\sim0.005$ days, but are valid over the entire temporal range if scaled according to the lightcurve evolution.

\subsection{Extinction in the host galaxy and intrinsic afterglow spectrum}
\label{alexsection}
Using the broadband UV-to-NIR SED derived in Sect. \ref{section:3}, we can derive the intrinsic host-galaxy extinction using the parametrisation of \cite{Pei1992ApJ} and following the method of e.g. \cite{Kann2006ApJ}. A fit without any extinction yields a very steep spectral slope $\beta_0=-2.90\pm0.03$ (usual intrinsic values range from $\approx-0.5-\,-1.1$), immediately indicative of dust along the line-of-sight in the host galaxy. The SED shows some scatter and strong curvature combined with small errors, leading to large $\chi^2$ values despite a visually good fit when a dust model is included. For Milky Way (MW), Large Magellanic Cloud (LMC), and Small Magellanic Cloud (SMC) dust, we derive positive intrinsic $\beta$ values, indicating it is unlikely that the host galaxy has dust similar to these local galaxies. Such positive values, with the flux density rising from the red to the blue, are not expected from afterglow theory. Of all three dust models, MW dust yields the best fit, with $\chi^2=1.36$, $\beta=0.32\pm0.14$, and $A_V=3.07\pm0.14$ mag. The two other dust models represent the SED less well (LMC: $\chi^2=3.61$, $\beta=0.61\pm0.14$, and $A_V=3.21\pm0.14$ mag; SMC: $\chi^2=5.91$, $\beta=0.39\pm0.13$, and $A_V=2.95\pm0.13$ mag), neither being able to adapt to the large $u-b$ colour and the relatively bright UVOT UV detections. Especially noteworthy is the failure of SMC dust, which is most often able to fit GRB sightlines well \citep[e.g.][]{Kann2006ApJ,Kann2010ApJ}. 

In addition to the three different fits with the intrinsic slope as a free parameter, we also fix the slope to two values based on the X-ray fit from the \emph{Swift} XRT archive, $\beta_X =\beta_{opt} = -0.81$ and $\beta_X-0.5 = \beta_{opt} = -0.31$. For these fits and MW dust, we derive $A_V=1.93\pm0.03$ mag and $A_V=2.41\pm0.04$ mag, respectively, indicating that $A_V = 1.9-2.4$ mag is a realistic range. Such large values of extinction had already been hinted at from spectroscopy \citep{2019GCN.23710....1K}, a result in qualitative agreement with the independent analysis of \cite{MagicMWLpaper}.

\begin{figure}
\centering
\includegraphics[width=1\columnwidth]{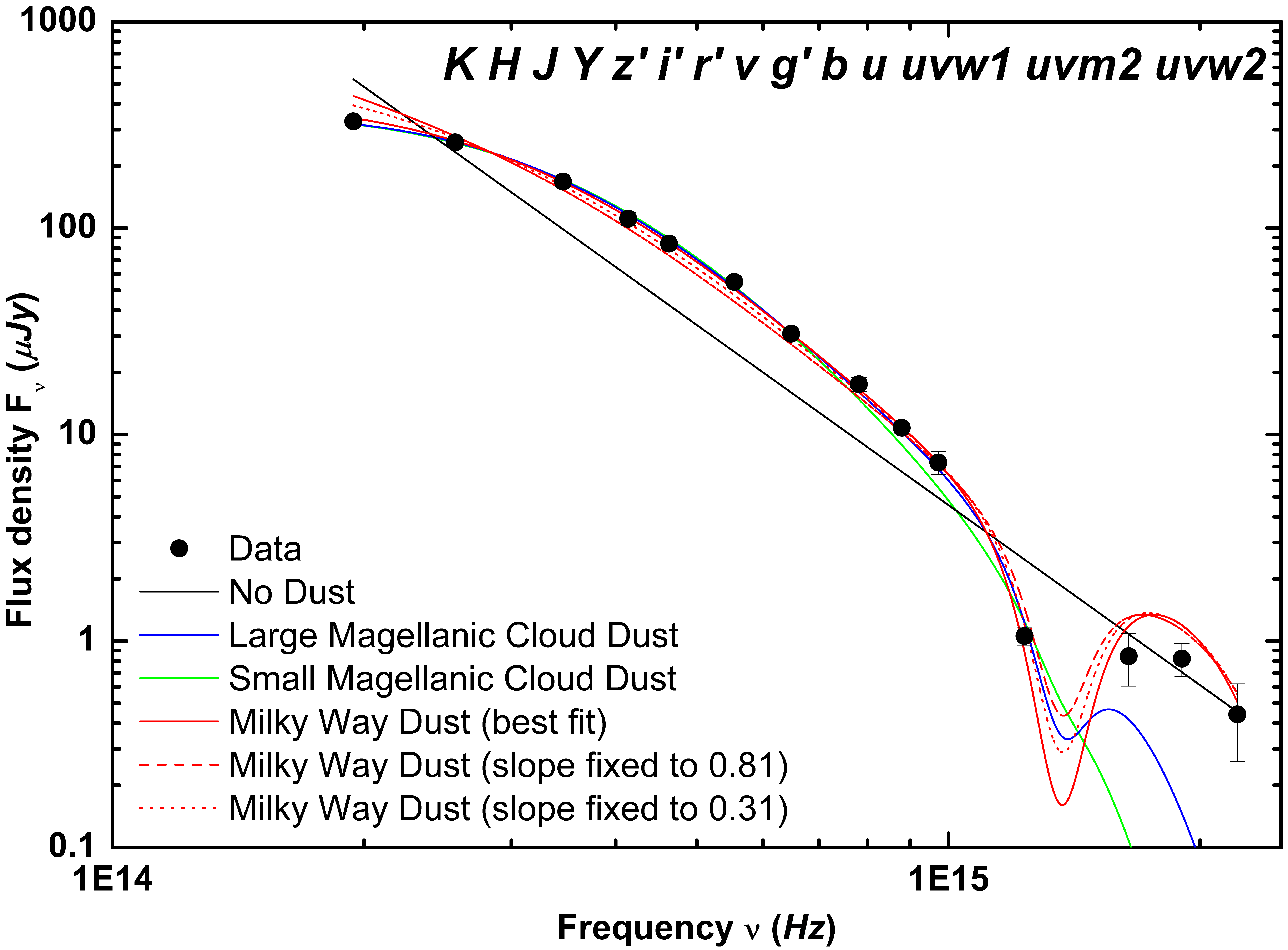}
\caption{Spectral Energy Distribution (SED) of GRB 190114C as derived from a multi-wavelength joint fit. It stretches from the UV (right) to the NIR (left). We have fit the SED with different dust-extinction laws. It can be seen that at $\lesssim10^{15}$ Hz, all three dust models fit about equally well, whereas there are large differences in the rest-frame far-UV; here, none of the dust models fits well, however, the Milky-Way dust fit fits the SED significantly better than the other two dust models. All fits indicate strong dust attenuation. For the MW fit, we also show two fits with fixed spectral slope $\beta$ derived from X-ray data (see text for more details).}
\label{fig:figureSED}
\end{figure}

The SED fits are shown in Fig. \ref{fig:figureSED}. It can be clearly seen that the three dust extinction laws differ little at $F_\nu\lesssim10^{15}$ Hz, implying that if low-$z$ bursts are only observed in the observer-frame $B$ band and redder, the dust model can not be determined \citep{Kann2006ApJ}. However, the detections in the UV clearly allow a distinction - and in this case, actually none of the three models fits the data well, however, the MW dust model yields the best of the three fits. While the detections in the three \emph{Swift} UVOT UV bands are low-S/N, they follow the afterglow decay as determined from the optical bands well, and the host galaxy is not luminous in these bands \citep{deUgartePostigo2019}. A more detailed analysis of the SED with a more free parametrisation than the curves of \cite{Pei1992ApJ} provide, following e.g. the methods of \cite{Zafar2018MNRAS,Zafar2018ApJL}, is beyond the scope of this paper.

\subsection{The afterglow of 190114C in the context of other GRB afterglows}
\label{section:31}
\begin{figure}
\centering
\includegraphics[width=1\columnwidth]{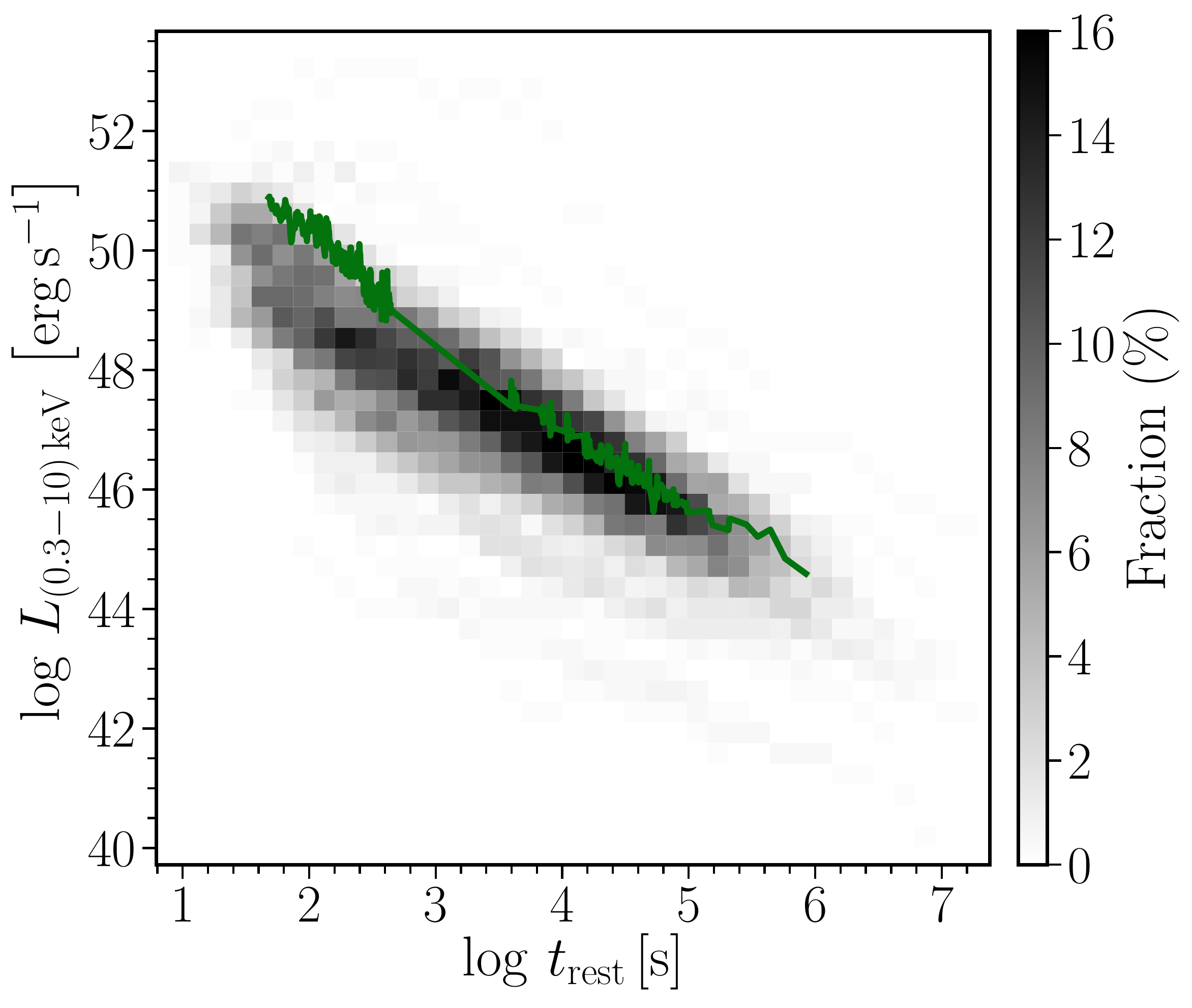}
\caption{The X-ray lightcurve of GRB 190114C in the context of the X-ray afterglows of \swift\ GRBs with known redshifts. The luminosity of GRB 190114C is comparable to the bulk of the GRB population.}
\label{fig:xray_comp}
\end{figure}

To put the X-ray emission in the context of other GRB afterglows, we retrieved the X-ray lightcurves of all \swift\ GRBs until the end of February 2019 with detected X-ray afterglows (detected in at least two epochs) and known redshifts from the \swift\ XRT archive.
The density plot in Fig. \ref{fig:xray_comp} displays the parameter space occupied by these 415 bursts (using the method described in \citealt{Schulze2014AA}). GRB 190114C, displayed in green, has a luminosity that is similar to the majority of the GRB population.

To compare the optical afterglow lightcurves of GRB 190114C with other GRB afterglows,  we follow the steps described below. We use the SED derived for the afterglow of GRB 190114C to shift the data of individual bands to the $R$ band, after subtracting the individual host-galaxy contributions,  and then clean this composite lightcurve of outliers. Hereby, we use only NIR data at $t>7$ days as this is expected to not be affected by the SN contribution as much.  We then use our knowledge of the redshift and the host-galaxy extinction with the method of \cite{Kann2006ApJ} to determine the magnitude shift $dRc$. This shift (together with the time shift determined from the redshift) moves the lightcurve in such a way as it would appear if the GRB occurred at $z=1$ in a completely transparent universe -- the host-galaxy extinction is corrected for. The time, however, is still given in the observer frame. Applied to a large sample, this allows for a direct luminosity comparison. For GRB 190114C, the high extinction and low redshift essentially cancel each other out.
For the two fits with MW dust coupled to the X-ray spectral slope, we find $dRc=-0.944^{+0.054}_{-0.055}$ mag for the high-extinction case ($\beta=-0.31$) and $dRc=-0.169^{+0.051}_{-0.053}$ mag for the low-extinction one ($\beta=-0.81$). 

\begin{figure*}
\centering
\includegraphics[width=1\columnwidth]{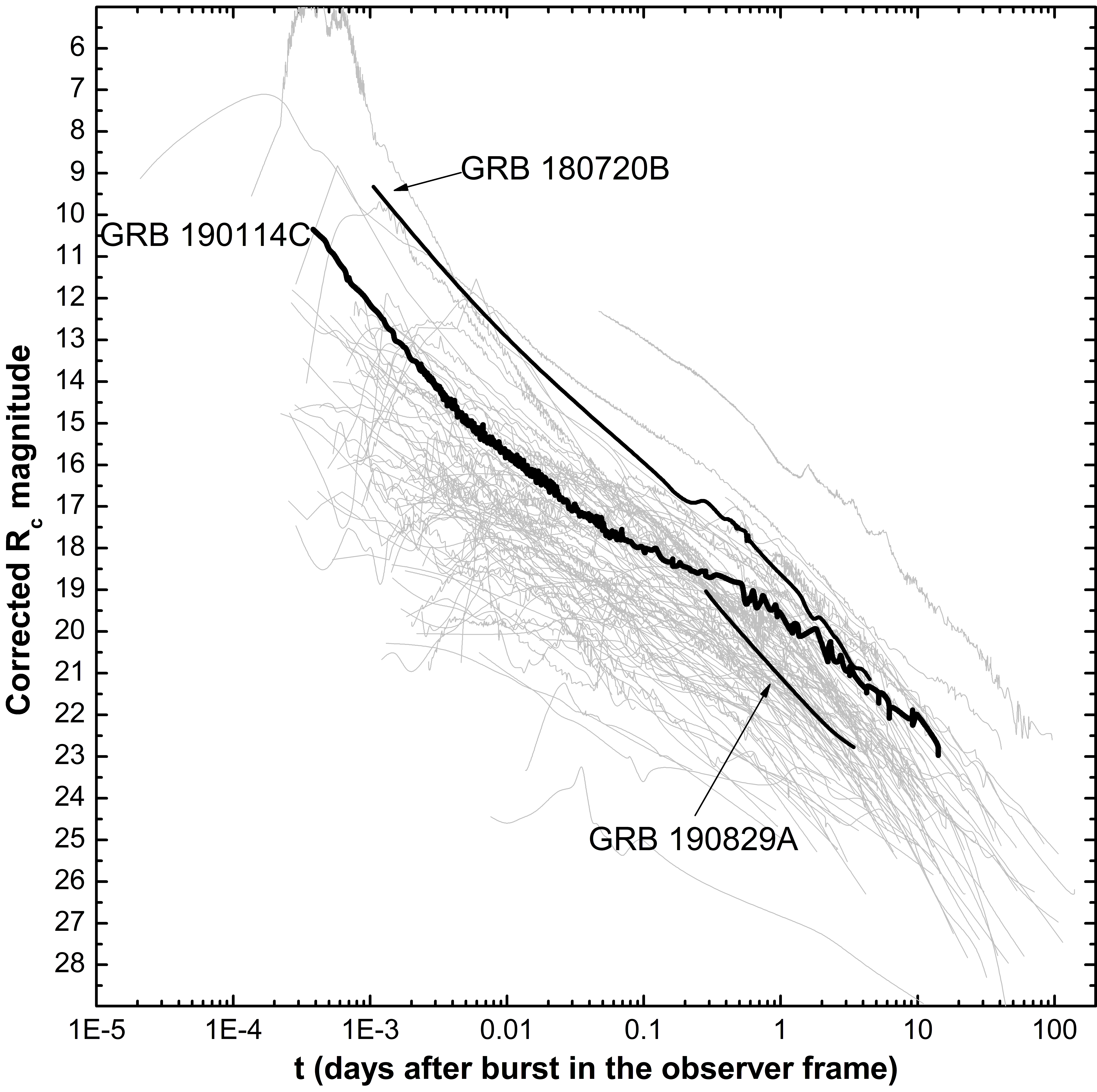}
\includegraphics[width=1\columnwidth]{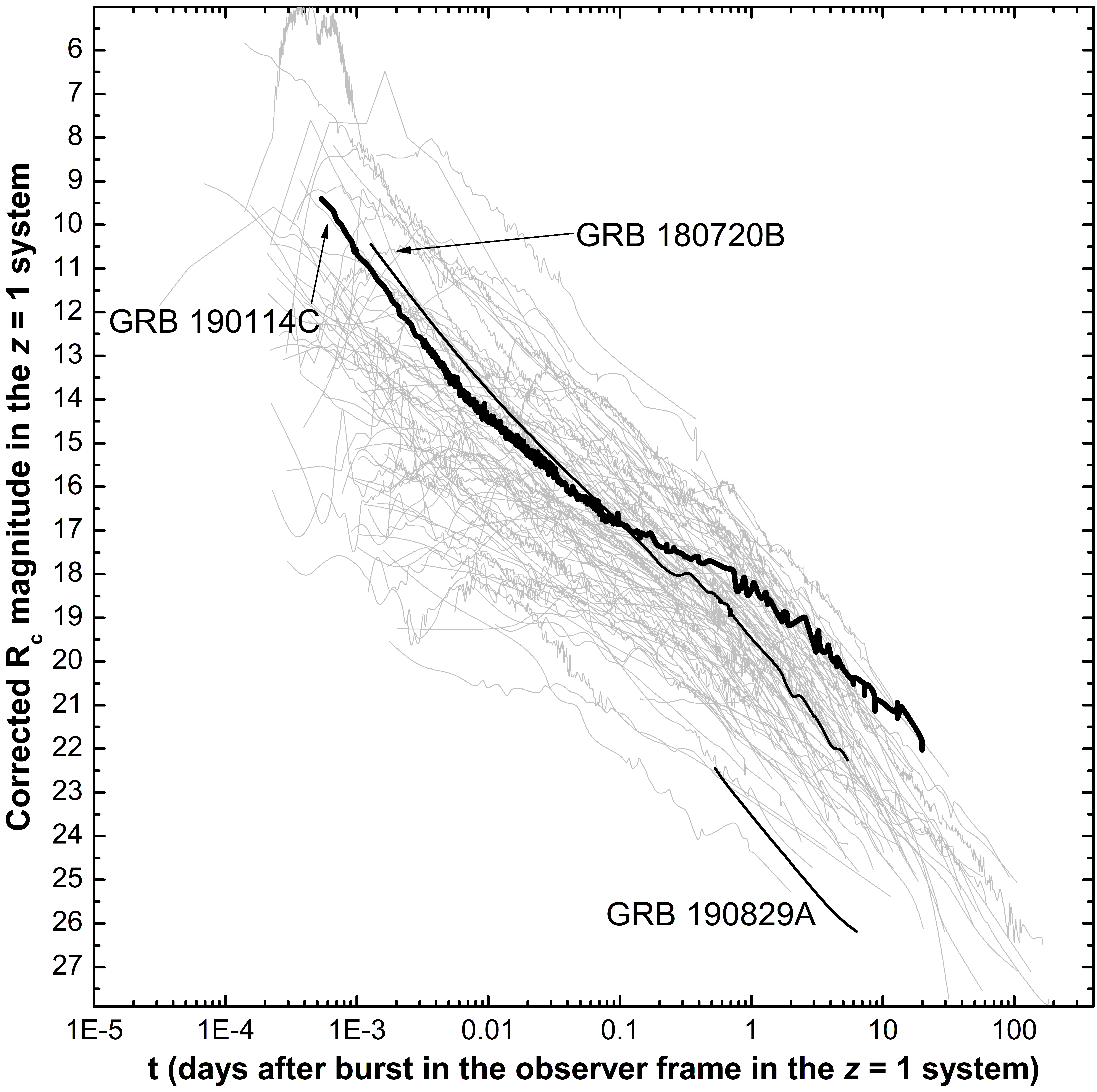}
\caption{The optical lightcurve of the afterglow of GRB 190114C in comparison to a large afterglow sample. \textbf{Left:} Afterglows in the observer frame. These have been corrected for Galactic extinction, as well as being host- and, where possible, supernova-subtracted, but are otherwise as observed. GRB 190114C is seen to have one of the brightest known early afterglows however, it is outshone by that of GRB 180720B. The third known GRB with a VHE detection, GRB 190829A, is seen to be of average brightness despite its proximity. \textbf{Right:} Afterglows shifted to the $z=1$ frame (see text for more details), corrected for host-galaxy extinction. Despite the large correction for extinction, the afterglow of GRB 190114C is seen to be of average nature at early times, but becomes one of the most luminous afterglows later on, stemming from the slow decay which is untypical for GRBs at $t>$ a few days. The other two VHE-detected GRBs are also plotted. The afterglow of GRB 180720B is seen to be similar to that of GRB 190114C, especially at early times. The afterglow of GRB 190829A, despite the correction for the large line-of-sight extinction, is seen to be underluminous, similar to other afterglows of nearby GRBs.}
\label{fig:figureKPs}
\end{figure*}

In Fig. \ref{fig:figureKPs}, we show the observed and corrected lightcurves of GRB 190114C in comparison to a large afterglow sample \citep{Kann2006ApJ,Kann2010ApJ,Kann2011ApJ,Kann2018AA}. The early steep decay likely resulting from a reverse-shock flash is clearly visible. At early times, the afterglow of GRB 190114C is one of the brightest detected so far observationally, despite the high line-of-sight extinction. However, in the $z=1$ frame (we plot the high-extinction case), it is seen to be of only average luminosity initially, making it once again similar to the ``nearby ordinary monster'' GRB 130427A \citep{Maselli2014Science}, and mirroring the result we find in the X-rays. At late times, the slow decay and lack of any visible jet break to $t>10$ days, which is unusual behaviour for GRB afterglows, leads it to become comparably more and more luminous. We caution this also stems from our choice of a high extinction correction, though, conversely, that is the better SED fit.
We can also compare the afterglow of GRB 190114C with the two other cases\footnote{Recently, GRB 201216C was also rapidly detected by MAGIC at VHE \citep{2020GCN.29075....1B}, and it shows clear evidence for very high line-of-sight extinction as well \citep{2020GCN.29077....1V}. The short GRB 160821B may have also been detected at VHE, albeit at low significance \citep{2021ApJ...908...90A}} of known GRB with VHE detections, namely GRB 180720B \citep{Abdalla2019Nature} and GRB 190829A \citep{deNaurois2019GCN25566}. Observationally, the afterglow of GRB 180720B (based on data from \citealt{Sasada2018GCN22977,Martone2018GCN22976,Horiuchi2018GCN23004,Itoh2018GCN22983,Reva2018GCN22979,Lipunov2018GCN23023,Schmalz2018GCN23020,Crouzet2018GCN22988,Zheng2018GCN23033,Watson2018GCN23017} as well as Kann et al. 2021a, in prep.) is seen to be even brighter, it is one of the few GRB afterglows detected at $<10$ mag at very early times. Despite its proximity, the afterglow of GRB 190829A (a preliminary analysis based on the data presented in \citealt{Hu2020AA}) is seen to be of average brightness, fainter than the more distant one of GRB 190114C.
We find a straight SED for the GRB 180720B afterglow, with no evidence for dust. This is very different from the two other VHE-detected GRBs, as we also find evidence for large line-of-sight extinction, $A_V\approx3$ mag, for GRB 190829A. Shifting both afterglows to $z=1$, we find the afterglow of GRB 180720B to be quite similar to that of GRB 190114C in terms of luminosity (albeit with a steeper decay at late times), and being an average afterglow in the context of the whole sample. The afterglow of GRB 190829A, on the other hand, despite the large extinction correction, is found to be less luminous than most of the sample, thereby being similar to those of other low-luminosity GRBs in the local Universe \citep{Kann2010ApJ}. The detection of VHE emission is therefore neither linked inextricably to the extinction along the line-of-sight, nor to the luminosity of the afterglow.

\begin{figure}
\centering
\includegraphics[width=1\columnwidth]{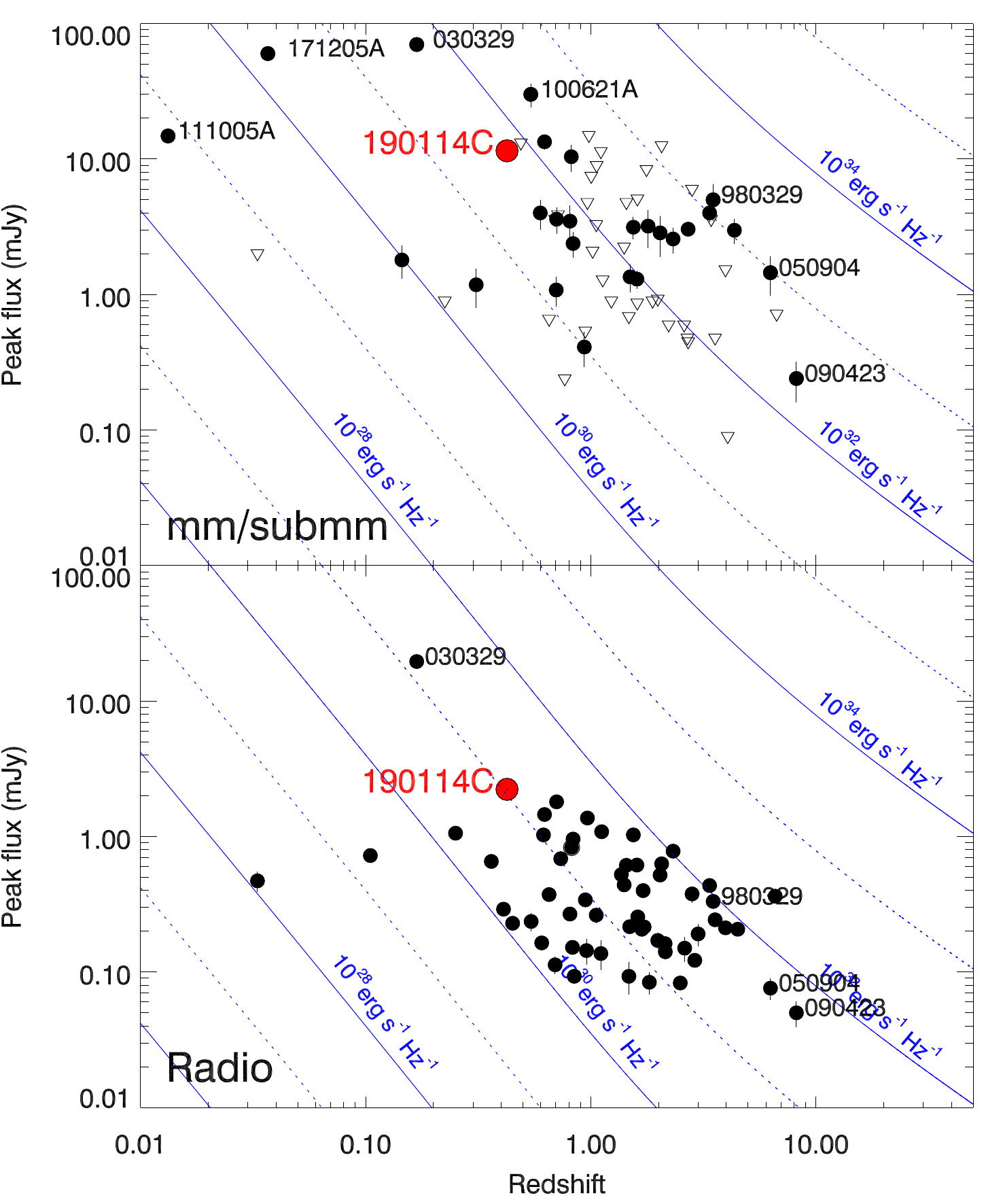}
\caption{Peak flux densities of GRB\,190114C in the context of the mm/submm and radio samples of \citet{deUgartePostigo2012} and \citet{Chandra2012}, plotted as a function of redshift. Blue lines indicate equal luminosity, with the most luminous events being found in the top right corner. Some prominent events have been highlighted.}
\label{fig:peakrad}
\end{figure}

Figure~\ref{fig:peakrad} shows a comparison of the peak flux densities of GRB 190114C with other events at millimetre and centimeter wavelengths, as a function of the redshift. Although GRB 190114C is bright at radio and millimetre wavelengths, this is mostly due to its low redshift. Comparing its peak luminosity with these samples of bursts, we yet again observe an average event. We note that the peak luminosity in millimetre wavelengths is dominated by the reverse shock, detected through very early ALMA observations by \citet{Laskar2019}.

\section{Interstellar Scintillation in Radio Bands}
\label{par:scint}
Inhomogeneities in the electron density distribution in the Milky Way along the GRB line-of-sight scatter radio photons. This effect, called interstellar scintillation (ISS), results in variations in measured flux density of the source at low frequencies ($\lesssim 10$~GHz, \citealt{Rickett90,Goodman97,Walker98,Goodman06,Granot14}). GRBs often display a similar behavior in their radio lightcurves (see e.g. \citealt{Goodman97,Frail97,Frail00}) with the variations occurring between observations on timescales ranging between hours and days.

In the standard (and easy) picture, ISS occurs at a single ``thin screen" at some intermediate distance $d_{scr}$, typically $\sim 1$~kpc for high Galactic latitudes.  The strength of the scattering is quantified by a dimensionless parameter, defined as \citep{Walker98,Walker01a}
\begin{equation}
\xi = 7.9 \times 10^3 SM^{0.6} d_{scr}^{0.5} \nu_{GHz}^{-1.7}
\label{eq:xi_u_iss}
\end{equation}
where $SM$ is the scattering measure (in units of kpc m$^{-20/3}$).

There are in general two types of ISS: weak and strong scattering. In particular, strong scattering can be divided into refractive and diffractive scintillation.
ISS depends strongly on the frequency: at high radio frequencies only modest flux variations are expected, while at low frequencies strong ISS effects are important. The transition frequency $\nu_{trans}$ between strong and weak ISS is defined as the frequency at which $\xi = 1$ \citep{Goodman97}:
\begin{equation}
\nu_{trans} = 10.4 \ SM_{-3.5}^{6/17} d_{scr}^{5/17} \ \mathrm{GHz}
\label{eq:nu_trans}
\end{equation}
where $SM_{-3.5} =$ ($SM/10^{-3.5}$ m$^{-20/3}$ kpc).
In the strong ISS regime, diffractive scintillation can produce large flux variations on timescales of minutes to hours but is only coherent across a bandwidth $\Delta \nu = (\nu / \nu_{trans})^{3.4}$ \citep{Goodman97,Walker98}. Refractive scintillation is broadband and varies more slowly, on timescales of hours to days.

In all regimes, the strength of scattering $\xi$ decreases with time at all frequencies as the size of the emitting region expands, with diffractive ISS quenching before refractive ISS. The source expansion also increases the typical timescale of the variations for both diffractive and refractive ISS (\citealt{Resmi17}). In this complex scenario, the contribution of ISS for each regime is defined by the modulation index $m$, defined as the rms fractional flux-density variation (e.g. \citealt{Walker98,Granot14}).

In our analysis we estimated the ISS effects on GRB\,190114C through a dedicated fitting function that includes both diffractive and refractive contributions \citep{Goodman06}. The values of $\nu_{trans} = 8.14$~GHz and $d_{scr} = 0.76$~kpc and $SM = 1.79 \times 10^{-4}$~kpc m$^{-20/3}$ are estimated through the NE2001\footnote{\url{http://www.astro.cornell.edu/\textasciitilde cordes/NE2001/}} model for the Galactic electron distribution \citep{Cordes02}. We estimated the ISS contribution in our radio data summing this effect to the uncertainty of flux densities; this contribution is very important in C (ATCA 5.5 GHz) and X (ATCA 9 GHz) bands ($\sim 50 \%$ of the flux density), whereas it is very low for L (GMRT 1.26 GHz) band and ALMA frequencies ($\lesssim 5\%$).

\section{X-ray, millimetre, and radio observations within the standard afterglow model}
\label{section:4}
We used the framework of the standard afterglow model (see \citealt{KumarZhang2014} for a review) to reproduce the multi-band afterglow evolution.  We primarily considered the radio/mm data presented in this paper for the modelling, along with the publicly available {\it Swift} XRT observations. We used the specific flux at $3$~keV for the model, obtained by converting the integrated flux in the $0.3-10$~keV band using an average spectral index of $-0.81$ quoted at the XRT spectral repository\footnote{https://www.swift.ac.uk/xrt\_spectra/00883832/ \citep{2009MNRAS.397.1177E}}. We excluded the optical/IR lightcurves because of the large host extinction (see section~\ref{section:31}), which introduces an additional parameter in the problem. We show optical predictions from parameters estimated by using X-ray and radio bands. 

The basic physical parameters of the afterglow fireball, isotropic equivalent energy $E_{\rm iso}$, ambient density ($n_0$ for ISM and $A_\star$ for wind), fractional energy content in electrons ($\epsilon_e$) and magnetic field ($\epsilon_B$) translate to the basic parameters of the synchrotron spectrum which are the characteristic frequency ($\nu_m$), cooling frequency ($\nu_c$), self-absorption frequency ($\nu_a$), and the flux normalization at the SED peak ($f_m$) at a given epoch \citep{WijersGalama99}. In addition, the model also depends on the electron energy spectral index $p$ and the fraction $\zeta_e$ of electrons going into the non-thermal pool. We use a uniform top-hat jet with half-opening angle $\theta_j$. 

We do not consider  synchrotron self-Comtpon (SSC) emission in our  model and hence we exclude MAGIC and \fermi\ LAT data from our analysis.

\subsection{A challenge to the standard model}
As mentioned in section \ref{section:3}, the XRT lightcurve decays with a slope of $\alpha_X=-1.344\pm0.003$ at $t \leq 10$~days and the ATCA lightcurves decay with a slope of $\alpha_{\rm radio} \sim -1$ at $t \geq 10$~days. The last XRT detection at $13.86$~days mildly deviates from the single power-law while the $3\sigma$ upper limit at $27.5$~days can not place any further constraints on a potential break. This may indicate the onset of jet effects at $\sim 10$~days, either due to a change in the dynamical regime or due to relativistic effects in case of a non-expanding jet \citep{Rhoads:1999wm, Sari:1999mr}. However, to begin with, we consider both lightcurve slopes to be devoid of jet effects (see section \ref{jetbreak} below for a discussion considering jet side effects). 

The difference $\Delta\alpha$ in the temporal indices of the two lightcurves is consistent with 0.25, the expected number if the synchrotron cooling break $\nu_c$ remains between the bands. Under this assumption, lightcurve slopes $\alpha_X$ and $\alpha_{\rm radio}$ imply $p \sim 2.45$ and a constant density ambient medium. However, this picture demands that the XRT spectral index should be $\sim -1.23$, which is not consistent with the value of $\beta_X=-0.81\pm0.1$ reported in the \swift\ XRT spectral repository. Moreover, if $\nu_c$ is between radio and X-ray frequencies, the spectral slope between radio (say $9$~GHz as a representative frequency) and XRT should lie between $-(p-1)/2 \sim -0.73 $ and $-p/2 \sim -1.2$, with the exact value decided by the position of $\nu_c$ at the epoch at which the spectral slope is measured.  To test the possibility of the X-ray lightcurve originating in the $\nu > \nu_c$ segment and the decaying part of the radio lightcurve belonging to the $\nu_m < \nu < \nu_c$ segment, we constructed a synthesised simultaneous spectrum at $10$~days, extrapolated from single power-law fits to the lightcurves at $5$~GHz, $9$~GHz, and $7.26 \times 10^{17}$~Hz ($3$~keV). We found that the ratio of the extrapolated fluxes is $F_X/F_{\rm 9GHz} = \left( \nu_X/\nu_{\rm 9GHz} \right)^{-0.64}$, which is even smaller than $\nu^{-0.73}$, completely ruling out the possibility of a $p \sim 2.45$.  

\begin{figure}
    \centering
    \includegraphics[scale=0.6]{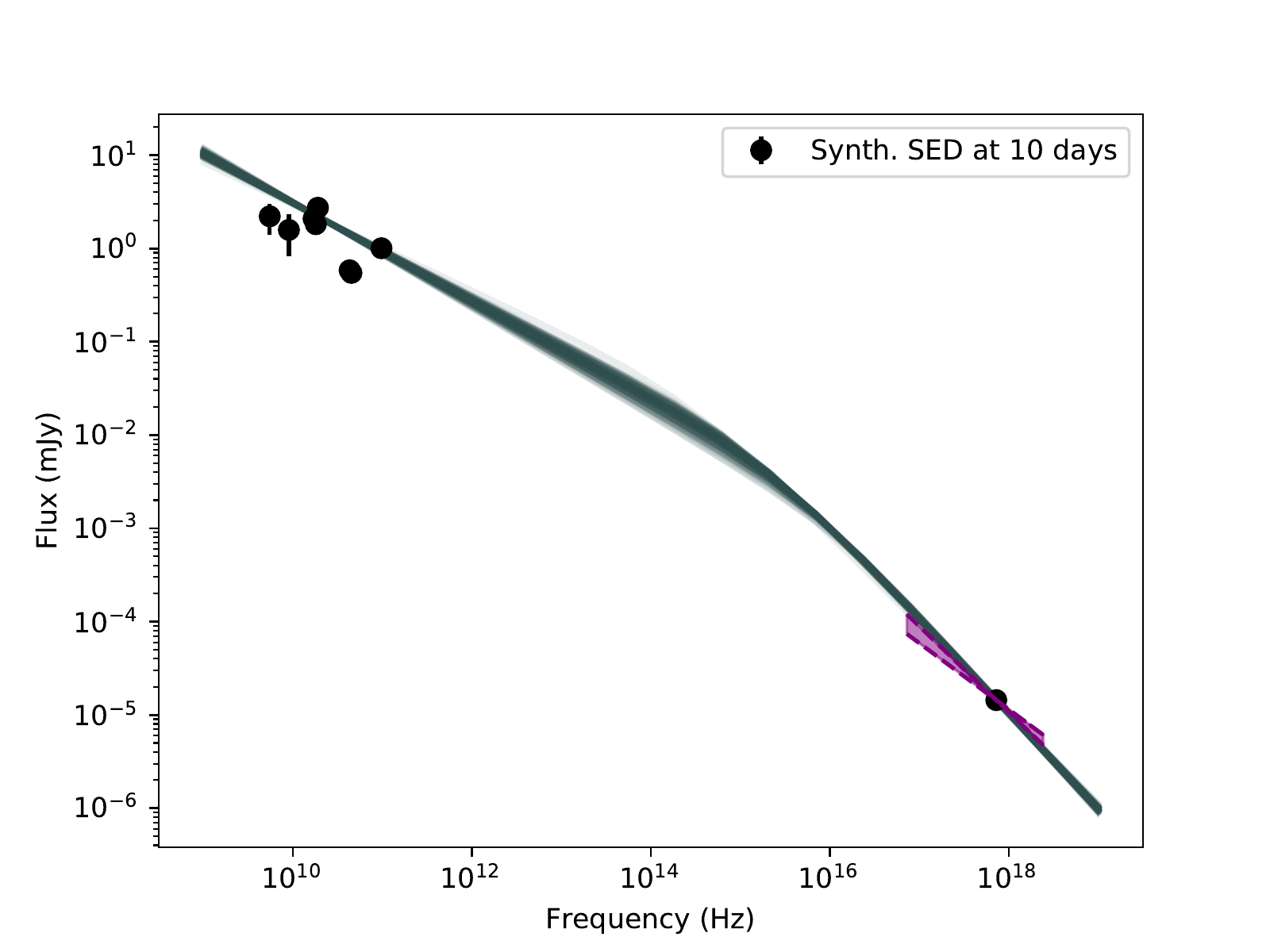}
    \caption{The synthesized radio-XRT spectrum at $\sim 10$~days. The smooth double power-law model assumes $\nu_c = 3.1 \times 10^{15}$~Hz and $\beta_1 = 0.52$. The observed spectral index from XRT spectral repository is shown in purple.}
    \label{fig10sed}
\end{figure}

Next we examine if the radio-XRT SED at $10$~days agrees with a smooth double power-law of asymptotic slopes $\beta_1$ and $\beta_2 = \beta_1+0.5$ (to mimic the synchrotron spectrum around $\nu_c$). We found the SED can be reproduced if $\beta_1 = -0.52\pm0.02$ and $\nu_c = 3.1^{+3.1}_{-1.5} \times 10^{15}$~Hz (Fig.~\ref{fig10sed}). The smoothing index is set at $2.0$. Due to the scatter in the radio/mm data at this epoch, we have used the scintillation correction described in section \ref{par:scint} to do the fitting. The fitting results are sensitive to the choice of data, such that without scintillation correction in $9$ and $5$~GHz, the spectral index is flatter.
Both the $\beta_1$ and the observed XRT spectral index are consistent with $p \sim 2.0$. Therefore, we conclude that while $p \sim 2$, both the radio and X-ray lightcurves decay at much steeper rate than expected, and the most likely solution is to relax the assumption that $\epsilon_e$ and $\epsilon_B$ are constants in time. It is to be noted that the best-fit host extinction correction leads to a flatter UV/optical/NIR SED. However, due to the uncertainties in inferring the host extinction (see section \ref{alexsection}), we have ignored this inconsistency in future discussions and also chose not to include UV/optical/NIR data for further analysis.

Nevertheless, in Appendix B we give a detailed description of how the radio/X-ray data compare with the standard afterglow model with constant $\epsilon_e$ and $\epsilon_B$. Before proceeding with the time-evolving microphysics model, we however explore the validity of a model with jet break at $\sim 10$~days in the next section. 

\subsubsection{Can a jet break save the standard model?}
\label{jetbreak}
The last XRT observation at $27.5$~days yielded an upper limit, which (within $3 \sigma$) falls above the extrapolation of the single power-law lightcurve. Yet, it is possible that there indeed is a break at $\sim 10$~days in the XRT lightcurve. More sensitive late-time observations by \textit{XMM-Newton} or \textit{Chandra} could yield conclusive evidence of this possibility. Considering the fact that the ATCA lightcurves also show a change of slope at about $10$~days, such a break can likely be due to jet effects, though achromaticity of jet breaks is debated \cite{Zhang2006a}. 

We consider two asymptotic examples, an exponentially expanding jet such as in \cite{Rhoads:1999wm} and a non-expanding jet. For the former, as the radial velocity is negligible post jet break, the temporal decay indices are insensitive to the density profile \citep{Rhoads:1999wm}. In this case, for the spectral regimes $\nu <\nu_a$, $\nu_a < \nu < \nu_m$, $\nu_m < \nu < \nu_a$, $(\nu_m, \nu_a) < \nu < \nu_c$, and $(\nu_m, \nu_c) < \nu$, the temporal indices are $0, -1/3, 1, -p$ and $-p$, respectively. However, the observed temporal decay of the ATCA lightcurves does not agree with any of these values, therefore this possibility is ruled out. Moreover, a smoothly varying double power-law fit to the XRT lightcurve (smoothing index of $2$) shows that the post-break slope $\alpha_{2,X} = -1.76 \pm 0.06$. This does not conform to the predictions of the simple model of exponentially expanding jets where the post break slope of the optically thin lightcurve is always $-p$.

For the latter case, the flux is reduced as the solid angle accessible to the observer increases beyond the jet edge. Therefore, the expression for the observed flux picks up an additional factor of $\Gamma^2$ (where $\Gamma$ is the bulk Lorentz factor of the jet) to account for the deficit in solid angle \citep{KumarZhang2014}. Here, for an adiabatic blast-wave in a constant density ambient medium ($\Gamma \propto t^{-3/8}$, \citealt{WijersGalama99}), post-break temporal indices are $-1/4$, $-1/4$, $+1/2$, $-3p/4$, and $-(3p+1)/4$, respectively for the above-mentioned set of spectral regimes. For a wind-blown density profile ($\Gamma \propto t^{-1/4}$, \citealt{Chevalier:1999mi}), the temporal indices become $+1/2$, $-1/2$, $+1/2$, $-(3p+1)/4$, and $-3p/4$, respectively. None of these values for a range of $2< p <3$ are in agreement with the radio lightcurve slope of $\sim -1$. Therefore, we rule out the possibility of a jet break saving the standard afterglow model. 

We conclude that even if there is an achromatic break in the lightcurves at $\sim 10$~days, non-standard effects are required to explain the multi-band flux evolution. In the next section, we describe the time-evolving shock micro-physics model. 

\subsection{Time-evolving shock micro-physics}
The standard afterglow model assumes that the fractional energy content in non-thermal electrons and the magnetic field, $\epsilon_e$ and $\epsilon_B$,  respectively, remain constant across the evolution of the shock. However, this need not necessarily be valid and there have been afterglows where micro-physical parameters have to be time-evolving \citep{Filgas2011b, 2014MNRAS.444.3151V}.
 
For simplicity, we consider a power-law evolution such as,
\begin{align}
\epsilon_e& \propto t^{\iota}, \nonumber \\
\epsilon_B & \propto t^{\lambda}.\nonumber
\end{align}
In such a model, if the general ambient medium density profile is characterised as $\rho(r) \propto r^{-s}$, the spectral parameters will evolve as,
\begin{align}
\nu_m \propto t^{\frac{(-3+4\iota+\lambda)}{2}}, \\
\nu_c  \propto t^{\frac{4+12\lambda-3s(1+\lambda)}{2(s-4)}}, \\
 f_m \propto t^{\frac{1}{2} \ (\lambda + \frac{s}{(s-4)})} , \\
\nu_a ( <\nu_m)  \propto t^{(\lambda- 5\iota+\frac{3s}{s-4})/5} , \\
\nu_a ( > \nu_m)  \propto t^{\frac{p(\lambda+4\iota-3)+2\left(\lambda-2\iota+\frac{s+4}{s-4}\right)}{2(4+p)}}.
\end{align}

To derive the equations governing the spectral parameters, we used the definitions given in \cite{WijersGalama99}. For optical depth to self-absorption, we used the expression given in \cite{Panaitescu2002a}, $\tau_m = \frac{5}{3-s} \frac{e n(r) r \zeta_e}{B \gamma_m^5}$, where $e$ is the electric charge, $n(r)$ is the ambient density as a function of the fireball radius $r$, $\zeta_e$ is the fraction of electrons in the power-law distribution, $B$ is the magnetic field, and $\gamma_m$ is the minimum Lorentz factor of the power-law electrons. For the dynamics of the blastwave in a constant density medium, we used the temporal evolution of bulk Lorentz factor and radius of the fireball given in \cite{WijersGalama99} and for the wind driven medium we used the same given in \cite{Chevalier:1999mi}. 

We first attempted $\iota=0$ (constant $\epsilon_e$). The lightcurve slope $\alpha_2$ for the spectral segment ($\nu > \nu_c$) then reduces to $(2+p(\lambda-3)-2\lambda)/4$ independent of the value of $s$, which equals the observed $\alpha_X$ only if $p>2.4$. Therefore, we conclude that time evolution of $\epsilon_B$ alone can not reproduce the observations.

We attempted Bayesian parameter estimation using Markov-Chain Monte-Carlo sampling under this model, but convergence could not be achieved perhaps due to the large dimension and degeneracy of the parameter space (see below). Therefore, we visually inspected the lightcurves to freeze the parameters which are sensitive to the lightcurve indices ($s, p, \lambda,$ and $\iota$).

\begin{figure}
    \centering
    \includegraphics[scale=0.25]{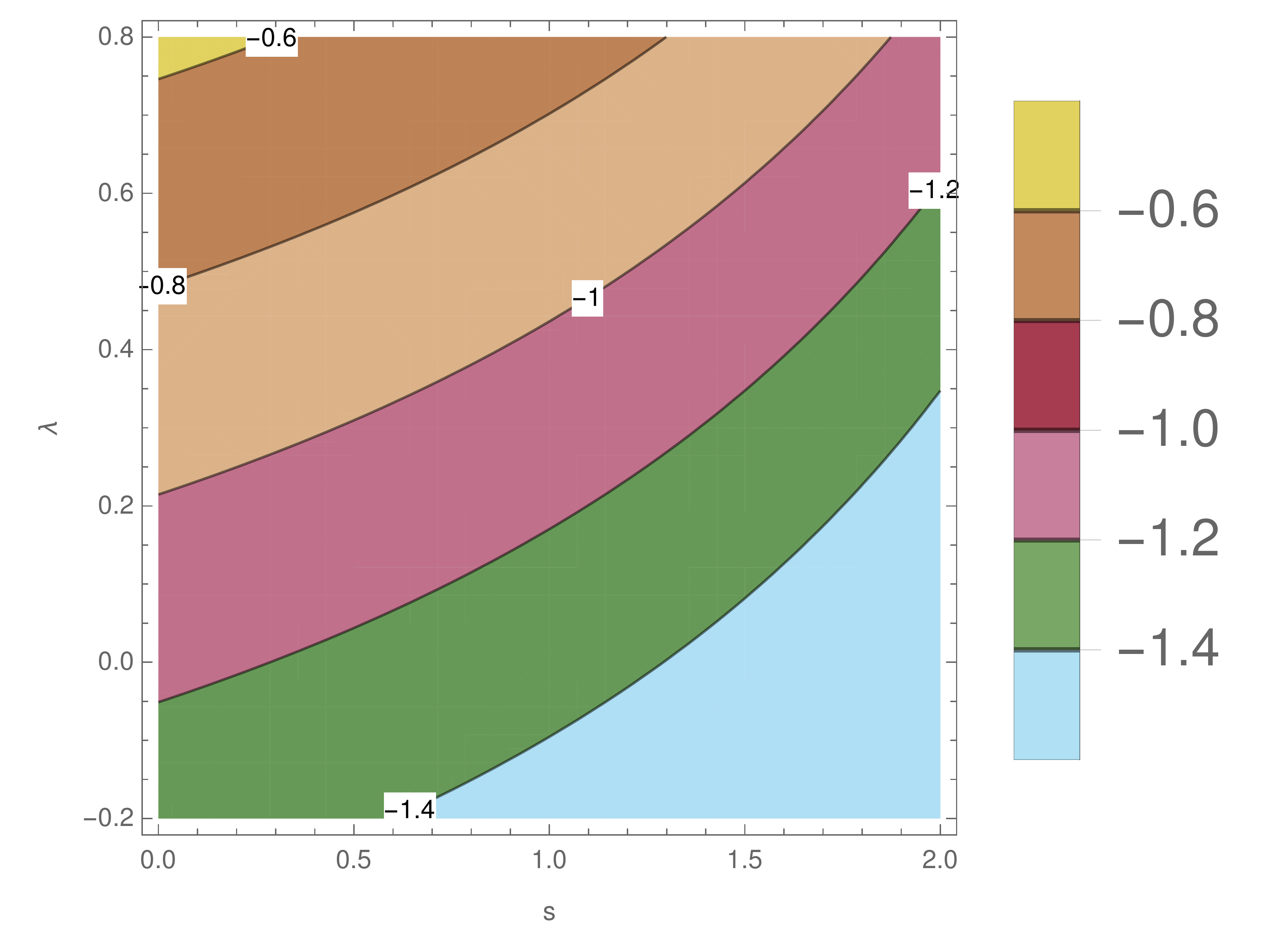}
    \caption{Predicted lightcurve decay slopes for the spectral segment $\nu_m < \nu< \nu_c$, as contours in the plane of $s-\lambda$. $p$ and $\iota$ are fixed at $2.01$ and $-0.4$ respectively. The observed radio decay index of $ -1.0$ to $-1.1$ can be reproduced by a range of $s$ and $\lambda$ values. In this paper, we have presented  models for ($s=0, \lambda=0.1$) and ($s=2, \lambda=0.76$).}
    \label{fig:slambdacntr}
\end{figure}

Using the results of the XRT, optical, and XRT/radio SED analysis, we fixed $p=2.01$. We used a value above $2$ to avoid the addition of yet another parameter to the problem, the upper cut-off of the electron distribution. When $p \sim 2$, $\alpha_2$ becomes a function of $\iota$ alone (dependence on $\lambda$ is weak for $p$ close to $2$ and zero for $p=2$) and we find that $\iota$ of $-0.4 - -0.3$ can reproduce the observed XRT lightcurve decay slope. For a fixed $p$ and $\iota$, a region of the $s-\lambda$ space can reproduce $\alpha_{\rm 9GHz}$ (see Fig.~\ref{fig:slambdacntr}). For a constant density medium, we fix $\lambda = 0.1$ and for a wind driven density profile, we fix $\lambda= 0.76$.

\begin{figure}
    \centering
    \includegraphics[width=\columnwidth]{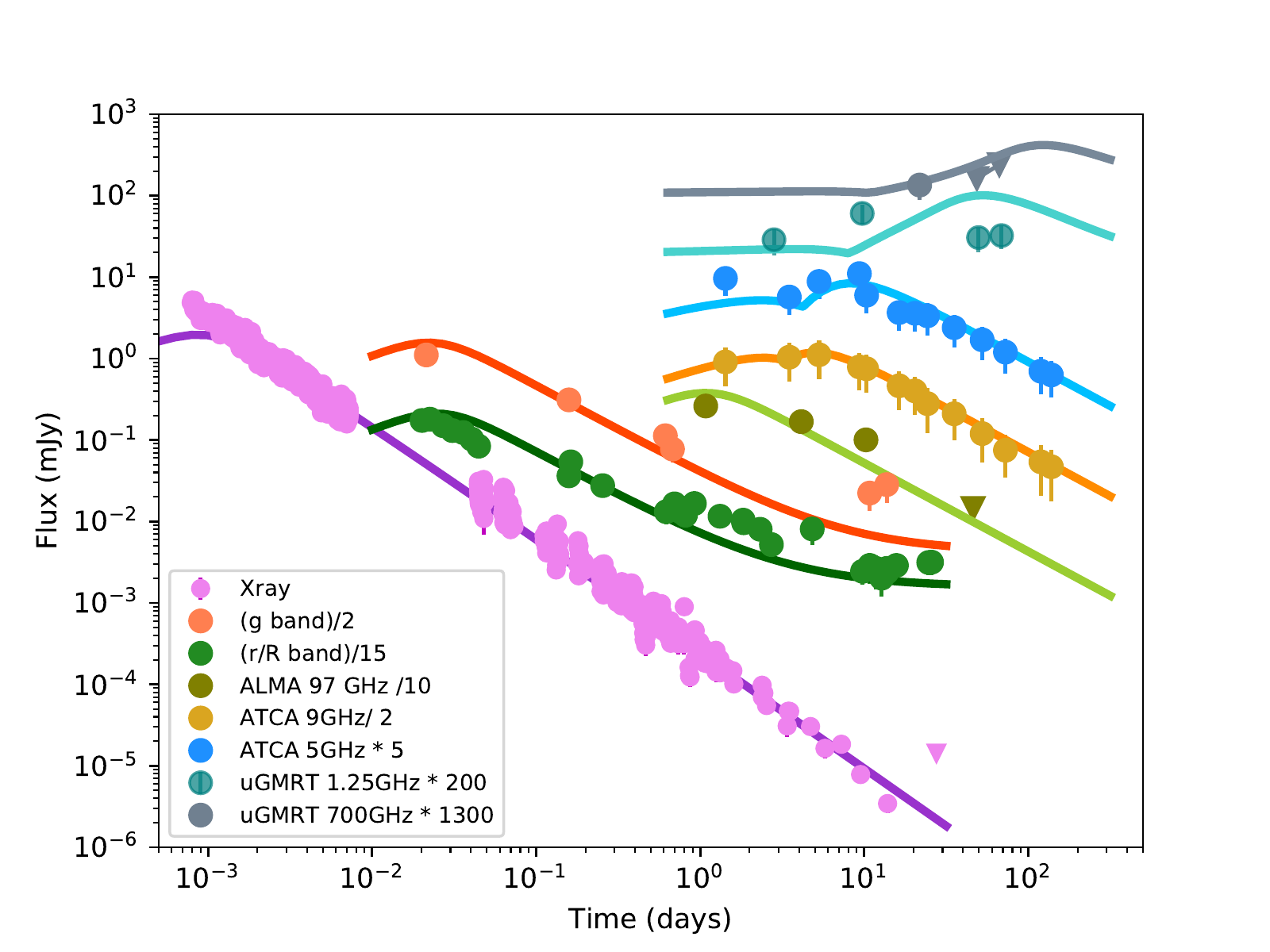}
    \caption{Lightcurves in the constant density model, corresponding to the peak of the posterior distribution presented in Fig. \ref{fig:posterior1}. Error bars in radio bands account for scintillation (see section \ref{par:scint}). Though we have included all the radio/mm data presented in this paper along with the XRT data in the Bayesian parameter estimation, for clarity we have only shown a few representative bands in this figure. For the uGMRT $1.25$~GHz ($0.65$~GHz) model, we have included a host galaxy flux of $0.05$~mJy ($0.07$~mJy). The optical bands are not included in the parameter estimation, but our model predictions, with the range of host galaxy extinction used in section- \ref{alexsection} ($A_v = 1.9 - 2.4$,  MW extinction law) are well in agreement with the observations. In this figure, we have used $A_v = 1.9$.}
    \label{fig:finalfig1}
\end{figure}

We attempted Bayesian parameter estimation in the spectral parameter regime, where the remaining parameters of the problem are $\nu_m$, $\nu_c$, $f_m$ and the optical depth $\tau_m$ at $\nu=\nu_m$. All values correspond to a specific epoch which we fixed to be $t=65$~s. We employed the Bayesian parameter estimation package pyMultinest \citep{2014A&A...564A.125B} based on the Nested Sampling Monte Carlo algorithm {\textit{Multinest}} \citep{2009MNRAS.398.1601F}. Multinest is an efficient Bayesian inference tool which also produces reliable estimates of the posterior distribution.

uGMRT measurements imply that the fireball is optically thick below $1.4$~GHz. However, as the low frequency data are limited, we could not obtain a meaningful convergence for $\tau_m$. Therefore, we ran simulations for different fixed values of $\tau_m$ and found that for the constant density medium, $-16.5 < \log \tau_m(t=65 {\rm sec}) < -15.5$ is consistent with the overall evolution of the fireball at higher frequencies. For the wind medium, $-17< \log \tau_m(t=65 {\rm sec}) <-12 $ is consistent. Nevertheless, we find that the self-absorbed lightcurves in $1.26$~GHz and $0.65$~GHz are not in great agreement with the observations. It is likely that the evolution of $\nu_a$ from this model is different from what is demanded by the observations (see Figs. \ref{fig:finalfig1} and \ref{fig:finalfig2}). A different $\nu_a$ evolution could arise due to absorption by thermal electrons (our solutions indicate a low fraction of electrons in the non-thermal pool, \citealt{Ressler:2017qjo}). A different $s-\lambda$ combination may also solve this discrepancy.

In Fig. \ref{fig:finalfig1} and Fig. \ref{fig:finalfig2} we present multi-band lightcurves from this model respectively for a constant density and a stratified density medium. For uGMRT $1.25$~GHz predictions, we have included a host galaxy flux of $0.05$~mJy (3 times the average RMS in our maps) to account for the host galaxy seen in meerKAT images \citep{Tremou2019GCN23760}. For uGMRT band-4 ($650$~MHz), we added a flux of $0.07$~mJy, considering a slope of $\nu^{-0.5}$ for the host SED. Even though the optical data are not included in the parameter estimation, we have presented $r^{\prime}$, $R$, and $g^{\prime}$ bands in the figure to demonstrate that both models are well in agreement with the early optical observations. At late time, optical transient flux includes contribution from the associated supernova which we have not considered in the model. In addition, the late time flux also contains  contribution from the host galaxy system, a close pair of interacting galaxies \citep{deUgartePostigo2019}. We have used host-galaxy magnitudes of $23.1$ (corresponding to HST F475W as the frequencies are close) in $g^{\prime}$ band. For $r^{\prime}$/R band, we used a host magnitude of $22$ which is in between that for F606W and F775W. It must be noted that depending on the telescope, the optical transient flux may also include contribution from the companion.

\begin{figure}
    \centering
    \includegraphics[width=\columnwidth]{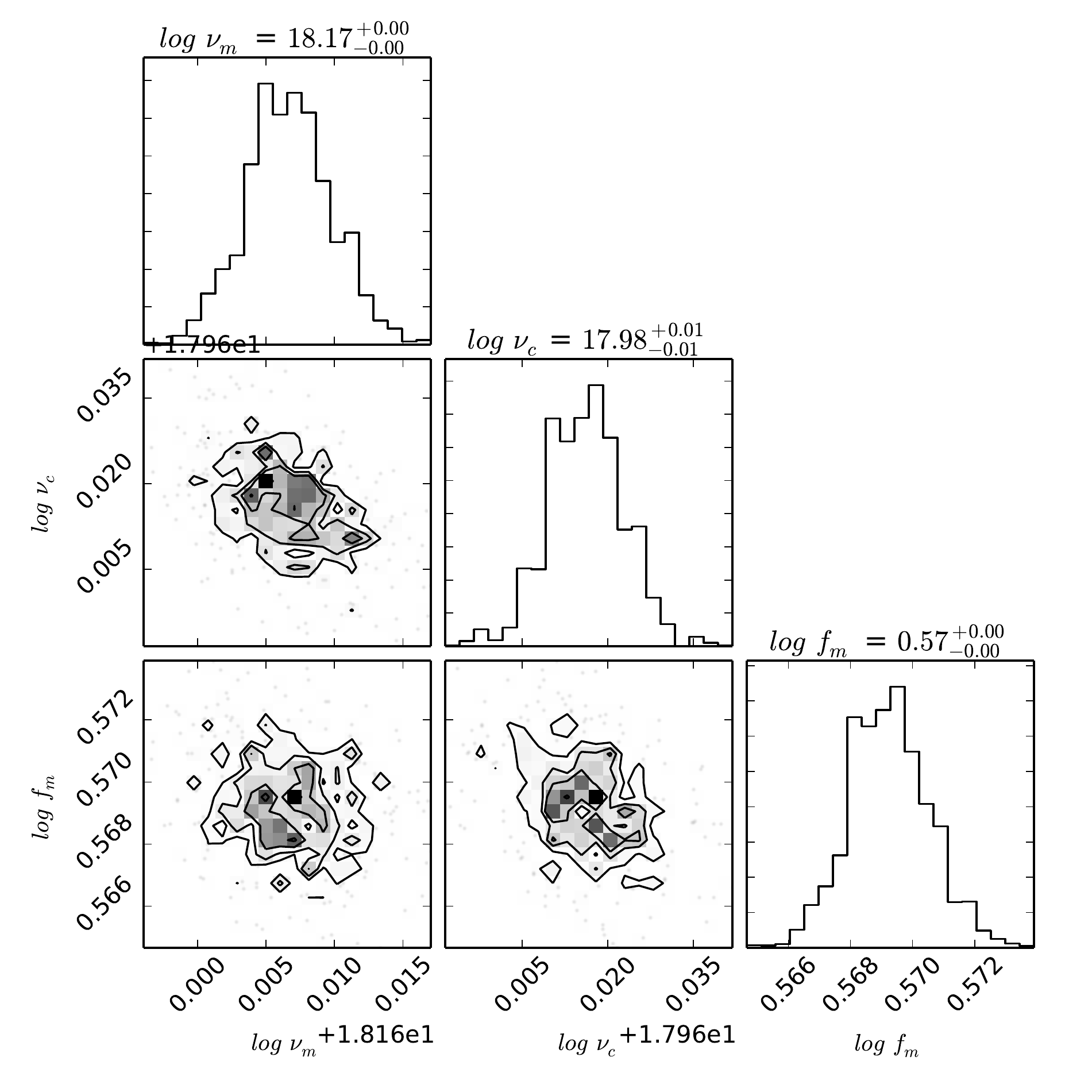}
    \caption{Posterior distributions of the three fit parameters, $\log(\nu_m/Hz), \log(\nu_c/Hz)$, and $log(f_m/mJy)$ for the constant density medium. Values correspond to $t=65$~s.}
    \label{fig:posterior1}
\end{figure}

\begin{figure}
    \centering
    \includegraphics[width=\columnwidth]{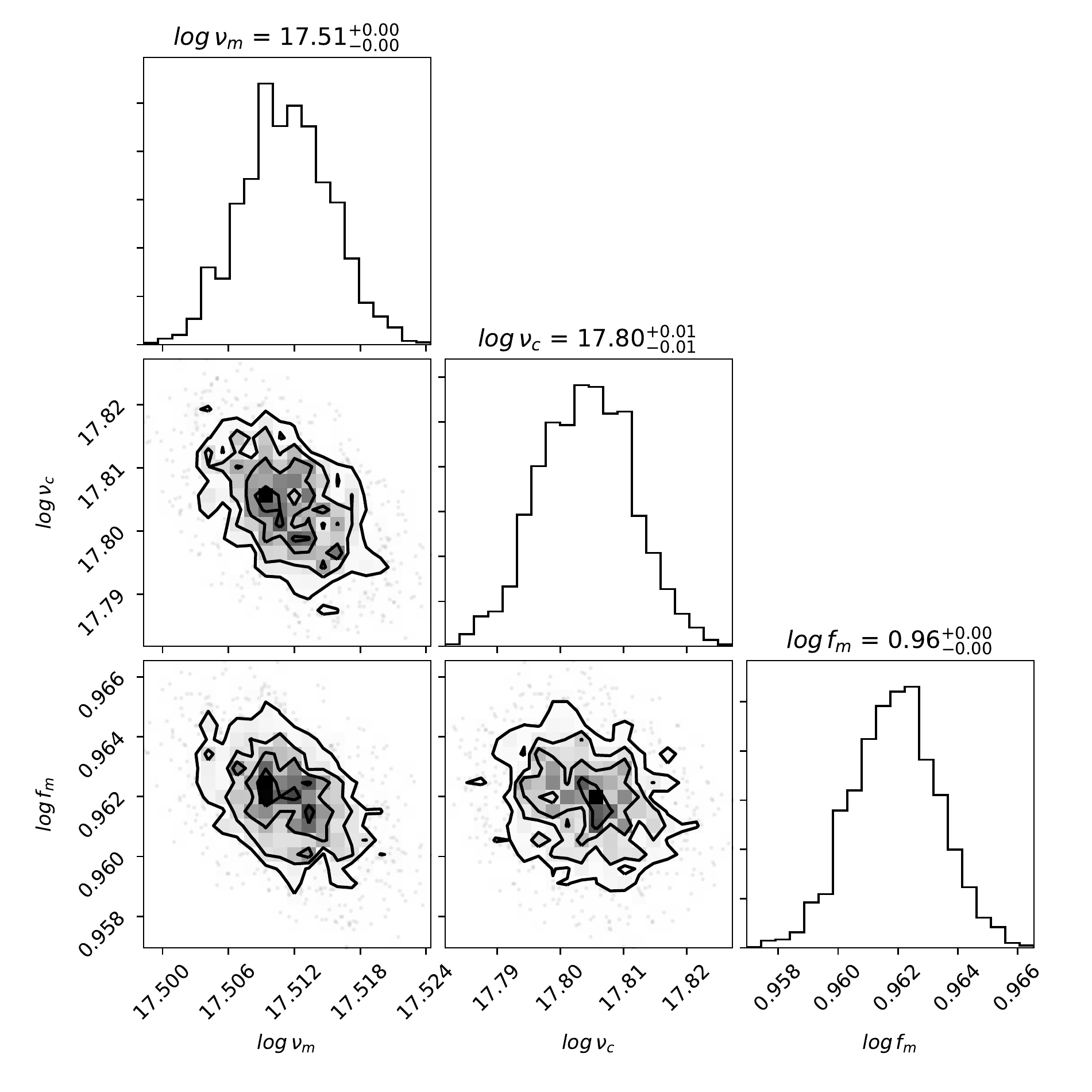}
    \caption{Posterior distributions of the wind model, same parameters as presented in Fig. \ref{fig:posterior1}}
    \label{fig:posterior2}
\end{figure}

\begin{figure}
    \centering
    \includegraphics[width=\columnwidth]{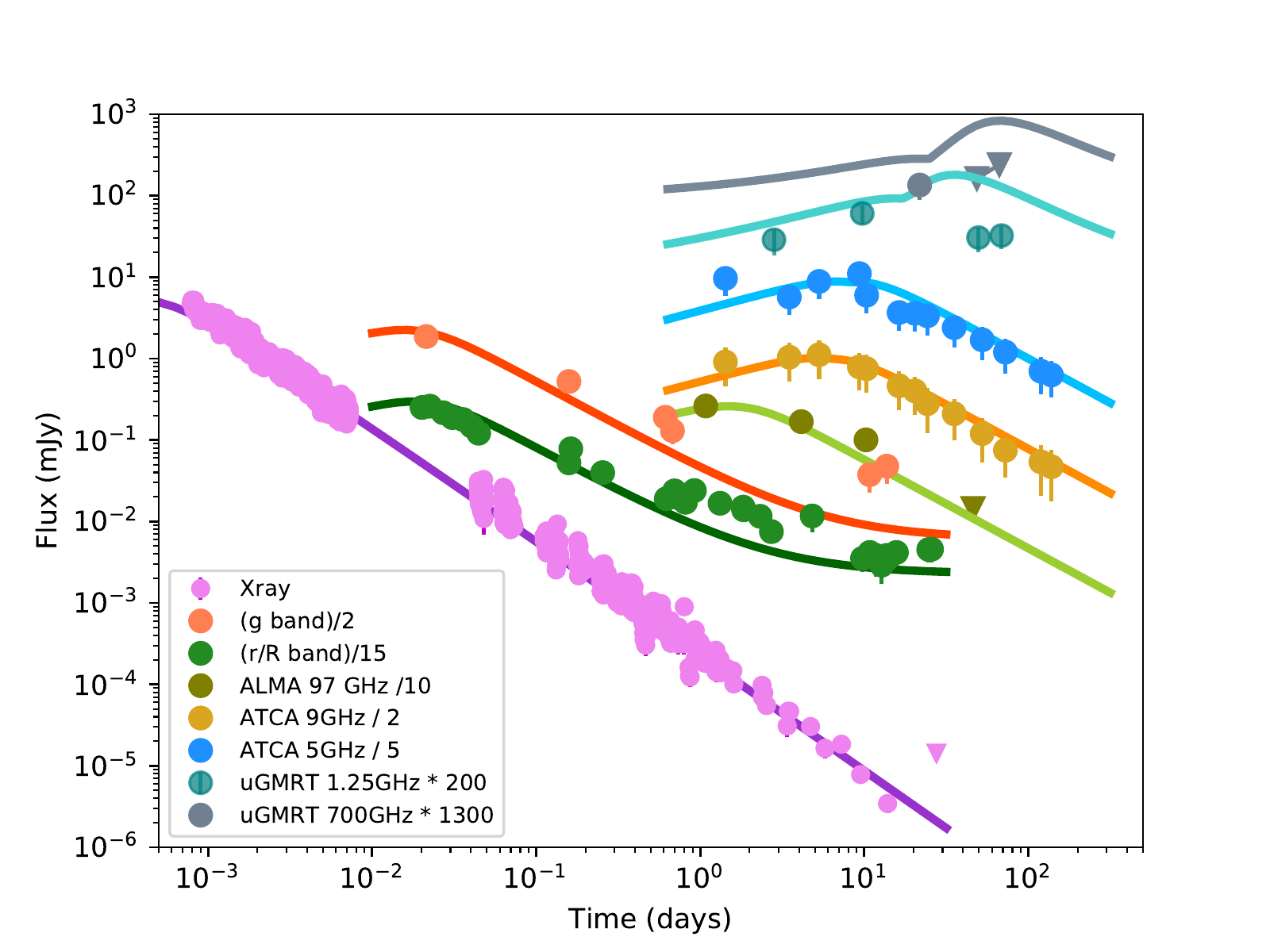}
    \caption{Lightcurves in the wind model, corresponding to the peak of the posterior distribution presented in Fig. \ref{fig:posterior2}. We have used host $A_v = 2.4$ for this figure.}
    \label{fig:finalfig2}
\end{figure}

In Fig. \ref{fig:posterior1} and \ref{fig:posterior2} we present the posterior distribution of the three-dimensional spectral parameter space and in tables \ref{tab:finaltab1} and \ref{tab:finaltab2}, we present the fit parameters and the inferred physical parameters. To derive the physical parameters we used the expressions given in Appendix A. The nested sampling global log-evidences are $-(0.37\pm0.44)$ and $-(0.38\pm0.25)$ for the ISM and the wind models respectively. As the values are comparable within errorbars, we can not prefer one ambient medium to the other under the premises of our model. 


\begin{table}
    \centering
    \begin{tabular}{c|c|c|c|c}
    \hline
    & \multicolumn{3}{c}{Fixed parameters} & \\
    %\hline \\
        $p$ & $\iota$ & $\lambda$&  $s$& $\log \tau_m$ \\
         $2.01$ &$-0.4$ &$+0.1$ &$0$ & $-16.0$ \\
    \hline \\
     & \multicolumn{3}{c}{Fitted parameters} & \\
      % \hline \\
     \multicolumn{2}{c}{$log \nu_m/Hz$}
     &\multicolumn{2}{c}{$log \nu_c/Hz$} & $log f_m/mJy$ \\
     \multicolumn{2}{c}{$18.167 \pm 0.003$}  &\multicolumn{2}{c}{$17.98 \pm 0.01$}  & $0.569 \pm 0.001$ \\
     \hline \\
     & \multicolumn{3}{c}{Derived physical parameters} & \\
    $E_{52}$ & $n_0$ & $\epsilon_{e0}$&  $\epsilon_{B0}$& $\zeta_e$ (assumed) \\
         $193.4$ &$23.0$ &$0.02$ &$4.7\times10^{-5}$ & $0.02$ \\
    \hline \\
    \end{tabular}
    \caption{Spectral and physical parameters of the time-varying micro-physics model. Spectral parameters are at $t=65$~sec. Physical parameters correspond to the peak of the posterior. Isotropic equivalent energy is normalised to $10^{52}$~ergs and number density is normalised to cm$^{-3}$. $\epsilon_{e0}$ and   $\epsilon_{B0}$ correspond to $t=1$~day. i.e, the final values of these parameters are $\epsilon_e = 0.02 (\frac{t}{\rm day})^{-0.4}$ and $\epsilon_B = 4.7 \times 10^{-5} (\frac{t}{\rm day})^{+0.1}$. The fraction of electrons $\zeta_e$ in non-thermal pool is decided by requiring $\epsilon_e (t=65)$~s $<1$.}
    \label{tab:finaltab1}
\end{table}

\begin{table}
    \centering
    \begin{tabular}{c|c|c|c|c}
    \hline
    & \multicolumn{3}{c}{Fixed parameters} & \\
    %\hline \\
        $p$ & $\iota$ & $\lambda$&  $s$& $\log \tau_m$ \\
         $2.01$ &$-0.4$ &$+0.76$ &$0$ & $-13.0$ \\
    \hline \\
     & \multicolumn{3}{c}{Fitted parameters} & \\
      % \hline \\
     \multicolumn{2}{c}{$log \nu_m/Hz$}
     &\multicolumn{2}{c}{$log \nu_c/Hz$} & $log f_m/mJy$ \\
     \multicolumn{2}{c}{$17.511 \pm 0.004$}  &\multicolumn{2}{c}{$17.804 \pm 0.007$}  & $0.961 \pm 0.001$ \\
     \hline \\
     & \multicolumn{3}{c}{Derived physical parameters} & \\
    $E_{52}$ & $A_\star$ & $\epsilon_{e0}$&  $\epsilon_{B0}$& $\zeta_e$ (assumed) \\
         $17.9$ &$2.0$ &$0.03$ &$4 \times10^{-4}$ & $0.02$ \\
    \hline \\
    \end{tabular}
    \caption{Spectral and physical parameters for the wind model. The final values of the microphysical parameters are $\epsilon_e = 0.02 (\frac{t}{\rm day})^{-0.4}$ and $\epsilon_B = 4.7 \times 10^{-5} (\frac{t}{\rm day})^{+0.76}$. The fraction of electrons $\zeta_e$ in non-thermal pool is decided by requiring $\epsilon_e (t=65)$~s $<1$.}
    \label{tab:finaltab2}
\end{table}

Our derived $E_{\rm iso}=1.93 \times 10^{54}$~ergs for the constant density medium exceeds the isotropic energy release in prompt emission by nearly an order of magnitude \citep{MagicMWLpaper}, while for the wind model ($E_{\rm iso} = 1.8 \times 10^{53}$~erg), it is nearly same as the prompt emission energetics. \cite{2004ApJ...613..477L} found that the prompt efficiency is nearly unity for bursts with large kinetic energy. The constant density solution leads this burst to have a much larger kinetic energy compared to bursts with similar prompt energy release, while by using the wind solution the burst lies close to the rest of the bursts in the distribution presented by \cite{2004ApJ...613..477L}. The $s-\lambda$ space is degenerate as we can see in Fig. \ref{fig:slambdacntr}. However, the extremely high value of $E_{\rm iso, K}$ may be used to disfavour the ISM medium.

Considering a lower limit at $t=100$~days for the jet-break, we derive the half opening angle of the jet to be $32.5^{\circ} \left( \frac{t_{\rm jet}}{100 \rm day} \right)^{3/8}$, indicating a true energy release of $E_{\rm tot} > 3 \times 10^{53}$~ergs for the constant density medium. For the wind medium, the opening angle from the same consideration of a jet break leads to $\theta > 19.3^{\circ}$, and a true energy release of $10^{52}$~ergs. The inferred steep rise in $\epsilon_B$ for a wind environment is still consistent with it to remain within the equipartition value within the available observations. The value of $\epsilon_B$ reaches $0.3$ only at $5000$~days.

\subsubsection{Reverse shock emission}
We used the $E_{\rm iso, 52}$ and the number density derived from the forward shock and searched the parameter space of the reverse shock (RS) to explain the early VLA and ALMA data presented by \cite{Laskar2019}. While the VLA data from $5-37$~GHz can be well explained by the RS model presented in \cite{ResmiZhang2016}, we could not reproduce the shallow decay of the $97$~GHz lightcurve around $0.1$~day. This could be resolved by improvements in the RS model. On the other hand, this may be resolved by a different combination of the degenerate parameter pair $s-\lambda$.

\subsection{Discussion on the modelling}
In summary, we have found that the multi-wavelength afterglow evolution is not consistent with the spectro-temporal closure relations predicted by the standard afterglow model. We have shown that a time evolution of the shock microphysical parameters can very well explain the overall behaviour of the afterglow, particularly above $1$~GHz. Such a time evolution of the afterglow shock microphysics has been invoked to explain individual afterglow observations in the past, for example by \cite{Filgas2011b} to explain GRB 091127 and by \cite{2014MNRAS.444.3151V} to explain GRB 130427A. For GRB 091127, an $\epsilon_B$ increasing as $t^{1/2}$ is required to explain the fast movement of the cooling break while for GRB 130427A, an $\epsilon_e \propto t^{-0.2}$ is required to explain the evolution of $\nu_m$. Compared to these authors, we require both $\epsilon_e$ and $\epsilon_B$ to evolve in time, and similar to our inferred evolution,  \cite{2014MNRAS.444.3151V} also require $\epsilon_e$ to decrease with time (though slower by a factor of 2). \cite{2014A&A...568A..45B} have invoked time evolution of shock microphysics in GRB prompt emission and have given a detailed description of the validity of this assumption in the context of Particle-in-cell (PIC) simulations of relativistic shocks.

{\cite{2018ApJ...866..162G} has found a scatter in the $p$ value between lightcurve and spectral indices for several GRBs. A standard deviation of $0.25$ found by them can accommodate the disagreement between the $\alpha$ and $\beta$ values for this burst. However, such a modelling incorporating a distribution of $p$ is beyond the scope of this paper.

We have not considered SSC cooling, which can modify the electron distribution and therefore cause deviation from the $\alpha-\beta$ closure relations expected under the standard model \citep{2009ApJ...703..675N}. However such a modification is expected only for the fast-cooling phase and it is highly unlikely to be relevant for late-time observations. 

It is also to be noted that numerical simulations of expanding jets have shown differences from semi-analytical treatments like ours (see \citealt{Granot14} for a review). For example, the jet break could be less pronounced in radio lightcurves. Therefore, employing results from more detailed numerical simulations may remove some of these inconsistencies. 

In addition to all these points above, another important fact to note is that the radio band is known to exhibit non-standard behaviour \citep{Frail:2003nv}, and the ATCA lightcurves may very well be representing the same. Detailed broadband follow-up of individual bursts in the radio band is definitely important and has promising prospects in the future era of the Square Kilometer Array and ngVLA.  

\section{Conclusions}
\label{conclusions}
In this paper we focus particularly on the late time and low frequency afterglow of the MAGIC-detected GRB 190114C obtained using GIT, DFOT and HCT in the optical, ALMA, ATCA and uGMRT in radio.  GRB 190114C is one of the three bursts (other two are GRBs 180720B and 190829A) so far detected at high GeV/TeV energies. Detailed modelling of the TeV and early multi-wavelength afterglow has shown that the high energy photons arise from up-scattered synchrotron photons \citep{MagicMWLpaper}. 

Multiwavelength evolution of the afterglow does not conform to the $\alpha-\beta$ closure relations expected under the standard fireball model. Mutiwavelength modelling indicates that for an adiabatic blastwave %{\sout{expanding into a constant-density ambient medium}},  
we require the microphysical parameters to evolve in time, as $\epsilon_e \propto t^{-0.4}$, $\epsilon_B \propto t^{+0.1}$ for constant density ambient medium and $\epsilon_B \propto t^{+0.76}$ for a stellar wind driven medium. %{\sout{However, this solution is not unique, and is valid for an assumed density profile.}} 
A time evolution of shock microphysics such as the one inferred here, resulting in a low $\epsilon_B$ and a high $\epsilon_e$ at early times may play a role in producing the bright TeV emission.  The inferred isotropic equivalent kinetic energy in the fireball, $1.9 \times 10^{54}$~ergs, exceeding that in the prompt emission as observed for several afterglows \citep{cenko2011}. Considering $100$~days as a lower limit to the jet break time, we derive the opening angle to be $> 32.5^{\circ}$ and the total energy to be $>3 \times 10^{53}$~ergs. 

Due to the inclusion of the late-time radio data, our interpretations differ from those of \cite{MagicMWLpaper} and \cite{Ajello:2019avs}. However, there are unsolved components in the evolution of the afterglow still, particularly in the early reverse shock emission at millimeter wavelengths. More detailed models including realistic jet dynamics and lateral expansion may have to be tested against these observations. These observations show the importance of low frequency campaigns in obtaining an exhaustive picture of GRB afterglow evolution.

\section*{Acknowledgements}
We thank the referee for providing constructive comments which have improved the scientific content of the paper. We thank the staff of the GMRT that made these observations possible. GMRT is run by the National Centre for Radio Astrophysics (NCRA) of the Tata Institute of Fundamental Research (TIFR). This paper makes use of the following ALMA data: ADS/JAO.ALMA\#2018.1.01410.T, ADS/JAO.ALMA\#2018.A.00020.T. ALMA is a partnership of ESO (representing its member states), NSF (USA) and NINS (Japan), together with NRC (Canada), MOST and ASIAA (Taiwan), and KASI (Republic of Korea), in cooperation with the Republic of Chile. The Joint ALMA Observatory is operated by ESO, AUI/NRAO and NAOJ.  The Australia Telescope Compact Array (ATCA) is part of the Australia Telescope National Facility which is funded by the Australian Government for operation as a National Facility managed by CSIRO.
The National Radio Astronomy Observatory is a facility of the National Science Foundation operated under cooperative agreement by Associated Universities, Inc. %VLA http://library.nrao.edu/ack.shtml
We thank Gaurav Waratkar, Viraj Karambelkar, and Shubham Srivastava for undertaking the optical observations with the GROWTH India Telescope (GIT). The GROWTH India Telescope (GIT) is a 70-cm telescope with a 0.7 degree field of view, set up by the Indian Institute of Astrophysics (IIA, Bengaluru) and the Indian Institute of Technology Bombay (IITB) with support from the Indo-US Science and Technology Forum (IUSSTF) and the Science and Engineering Research Board (SERB) of the Department of Science and Technology (DST), Government of India (https://sites.google.com/view/growthindia/). It is located at the Indian Astronomical Observatory (Hanle), operated by the Indian Institute of Astrophysics (IIA).  This work made use of data supplied by the UK Swift Science Data Centre at the University of Leicester.  L. Resmi and V. Jaiswal acknowledge support from the grant EMR/2016/007127 from Dept. of Science and Technology, India. D. A. Kann, A. de Ugarte Postigo, and C. Th\"one acknowledge support from the Spanish research project AYA2017-89384-P. AdUP and CT acknowledge support from funding associated to Ram\'on y Cajal fellowships (RyC-2012-09975 and RyC-2012-09984). D. A. Kann also acknowledges support from the Spanish research project RTI2018-098104-J-I00 (GRBPhot). KM, SBP and RG acknowledge BRICS grant DST/IMRCD/BRICS/Pilotcall/ProFCheap/2017(G) for this work.
V. Jaiswal and S. V. Cherukuri thank Ishwara-Chandra C. H. for kindly making GMRT data analysis pipeline available. L Resmi thanks Johannes Buchner for helpful discussions on pyMultinest. Harsh Kumar thanks the LSSTC Data Science Fellowship Program, which is funded by LSSTC, NSF Cybertraining Grant \#1829740, the Brinson Foundation, and the Moore Foundation; his participation in the program has benefited this work. The Cosmic Dawn Center is funded by the DNRF.  JPUF thanks the Carlsberg Foundation for support. MJM acknowledges the support of the National Science Centre, Poland through the SONATA BIS grant 2018/30/E/ST9/00208. GEA is the recipient of an Australian Research Council Discovery Early Career Researcher Award (project number DE180100346) funded by the Australian Government.

%%%%%%%%%%%%%%%%%%%%%%%%%%%%%%%%%%%%%%%%%%%%%%%%%%

\section*{Data Availability}
The optical, radio and millimeter data underlying this article is available in the article.



%\bibliographystyle{mnras}
%\bibliography{bibliographys} % if your bibtex file is called example.bib

%\begin{thebibliography}{99}
%\end{thebibliography}

%%%%%%%%%%%%%%%%%%%%%%%%%%%%%%%%%%%%%%%%%%%%%%%%%%

%%%%%%%%%%%%%%%%% APPENDICES %%%%%%%%%%%%%%%%%%%%%
\appendix
\section{Spectral parameters for a model with time-evolving shock micro-physics}
In this section, we present the expressions for the spectral parameters used in the time-evolving microphysics model. For the adiabatic dynamics of the relativistic blastwave in a constant density medium, we used the expressions from \cite{WijersGalama99}. For the wind medium, we used the dynamics given in \cite{Chevalier:1999mi}. In these expressions, $E_{52}$ is the isotropic kinetic energy normalized to $10^{52}$~ergs, $n_0$ is the ambient medium density in units of cm$^{-3}$,  $A_\star$ represents the normalized density in a stratified wind medium following \cite{Chevalier:1999mi}, $\epsilon_{e0}$ and $\epsilon_{B0}$ are the fractional energy in electrons and magnetic field respectively at one day since explosion, $p$ is the electron energy spectral index, $\zeta_e$ is the fraction of electrons in the non-thermal pool, $X$ is the hydrogen mass fraction, $x_p$ and $\phi_p$ are numerical functions of $p$ as explained in \citet{WijersGalama99}, and $t_d$ is time since burst in days. $t_0$ is the reference time where $\epsilon_{e0}$ and $\epsilon_{B0}$ are measured.

Expressions of spectral parameters in the constant density medium are given below. 
\begin{align}
    \nu_m = \frac{6.83 \times 10^{16} {\rm Hz}\,  (p-2)^2\, x_p\, {E_{52}^{1/2}}\, {\epsilon_{B0}}^{1/2}\,  {\epsilon_{e0}}^2}{{(p-1)^2}\, (1+X)^2\, \zeta_e^2}\, t_d^{-3/2} \, \left( \frac{t_d}{t_0} \right)^{\lambda/2+2\iota}
\end{align}

\begin{align}
    \nu_c = \frac{9.42 \times 10^{11} {\rm Hz}}{{E_{52}^{1/2}}\, n_0\, {\epsilon_{B0}}^{3/2}}\, t_d^{-1/2} \, \left(\frac{t_d}{t_0}\right)^{-3\lambda/2}
\end{align}

\begin{align}
    f_m = 20.7{\rm mJy} \phi_p\,  (1 + X)\,  E_{52}\, n_0^{1/2}\, {\epsilon_{B0}}^{1/2}\, \zeta_e \left(\frac{t_d}{t_0}\right)^{\lambda/2}
\end{align}

\begin{align}
    \tau_m = \frac{3.38 \times 10^{-15}\, n_0\, (p-1)^5\, (1+X)^5\, \zeta_e^6}{{(p-2)^5\, E_{52}^{1/2}}\,  {\epsilon_{B0}}^{1/2}}\, {\epsilon_{e0}^5}\, t_d^{5/2} \left(\frac{t_d}{t_0}\right)^{(\frac{5 \iota}{2} -\frac{\lambda}{2})}
\end{align}

For the wind driven density profile, the expressions are, 
\begin{align}
    \nu_m = \frac{7.45 \times 10^{16} {\rm Hz}\,  (p-2)^2\, x_p\, {E_{52}^{1/2}}\, {\epsilon_{B0}}^{1/2}\,  {\epsilon_{e0}}^2}{{(p-1)^2}\, (1+X)^2\, \zeta_e^2}\, t_d^{-3/2} \, \left( \frac{t_d}{t_0} \right)^{(\lambda/2+2\iota)}
\end{align}

\begin{align}
    \nu_c = \frac{4.9 \times 10^{10} {\rm Hz}}{{E_{52}^{1/2}}\, A_\star^{2} \, {\epsilon_{B0}}^{3/2}}\, t_d^{-1/2} \, \left(\frac{t_d}{t_0}\right)^{-3\lambda/2}
\end{align}

\begin{align}
    f_m = 564.2{\rm mJy} \phi_p\,  (1 + X)\,  E_{52}^{1/2} \, A_\star \, {\epsilon_{B0}}^{1/2}\, \zeta_e t_d^{-1/2} \, \left(\frac{t_d}{t_0}\right)^{\lambda/2}
\end{align}

\begin{align}
    \tau_m = \frac{3.38467 \times 10^{-15}\, n_0\, (p-1)^5\, (1+X)^5\, \zeta_e^6}{{(p-2)^5\, E_{52}^{1/2}}\,  {\epsilon_{B0}}^{1/2}}\, {\epsilon_{e0}^5}\, t_d^{3/2} \left(\frac{t_d}{t_0}\right)^{(-5\iota -\frac{\lambda}{2})}
\end{align}

\section{Standard model fits}
In this section we present results from our attempts to test the standard model predictions with the XRT/radio/mm data. This is an illustration that the multi-wavelength data can not be explained by the model. We used both a constant density and wind-blown ambient medium for our analysis. 

Reverse shock emission depends, in addition to $E_{\rm iso}$, $\theta_j$, and the ambient density, on the initial bulk Lorentz factor $\eta$ of the fireball, the electron energy spectrum (characterised by $p_{\rm RS}$) and the fractional energy content in electrons and the magnetic field, parametrised as $\epsilon_e^{\rm RS} = {\mathcal R}_e \epsilon_e^{\rm FS}$ and $\epsilon_B^{\rm RS} = {\mathcal R}_B \epsilon_B^{\rm FS}$ respectively. 

The code used in this analysis accounts for forward and reverse (thick shell RS in a wind medium and thin shell RS in a constant-density medium) shock emission, and was developed in \cite{ResmiZhang2016}. We used the Bayesian parameter estimation package pyMultinest \citep{2014A&A...564A.125B, 2009MNRAS.398.1601F} to explore the seven-dimensional parameter space of $(p_{\rm FS}, \epsilon_e, \epsilon_B, \theta_J, E_{\rm iso,52}, n_0 \mathrm{or} A_\star, \eta)$. We fixed ${\mathcal R}_B=1$ and $p_{\rm RS}=2.1$, as keeping them as free parameters will not give any additional advantage in explaining this data. If X-ray or radio/mm is considered alone, excellent agreement with the model is possible. However, for both types of ambient medium, models fail to reproduce the afterglow evolution if the entire data set is considered.

\begin{figure*}
    \centering
    \includegraphics[scale=0.28]{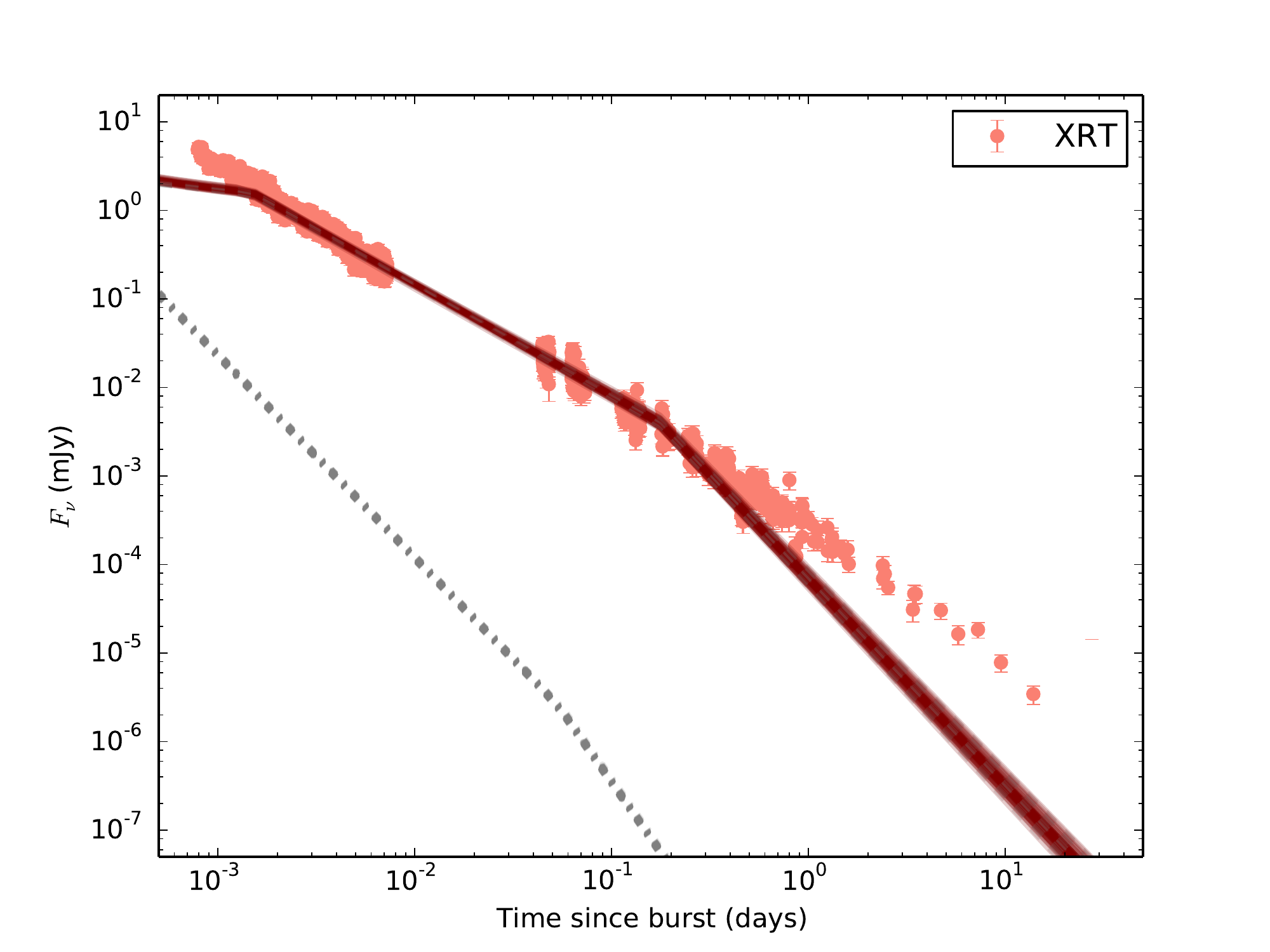}
    \includegraphics[scale=0.28]{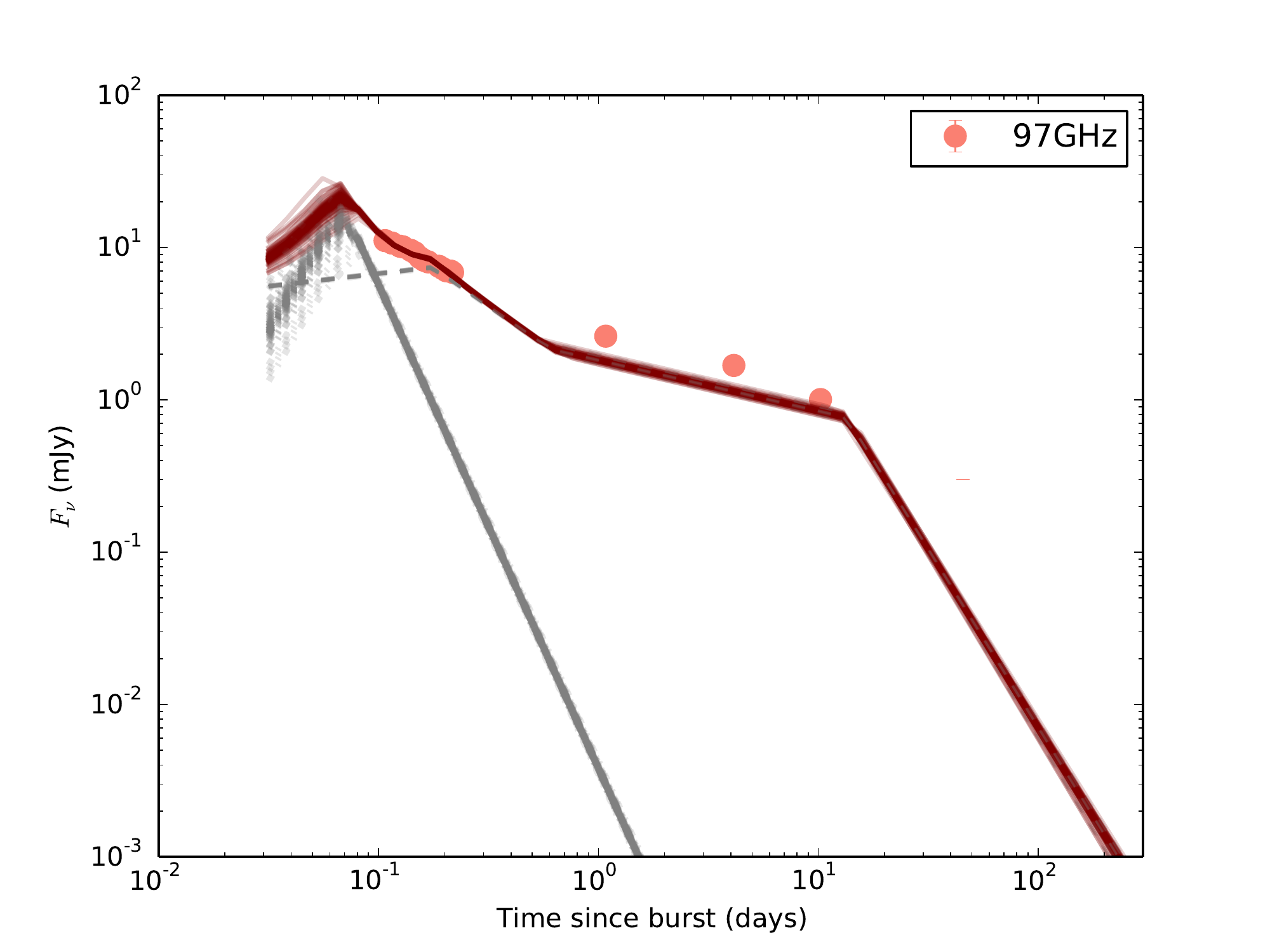}
    \includegraphics[scale=0.28]{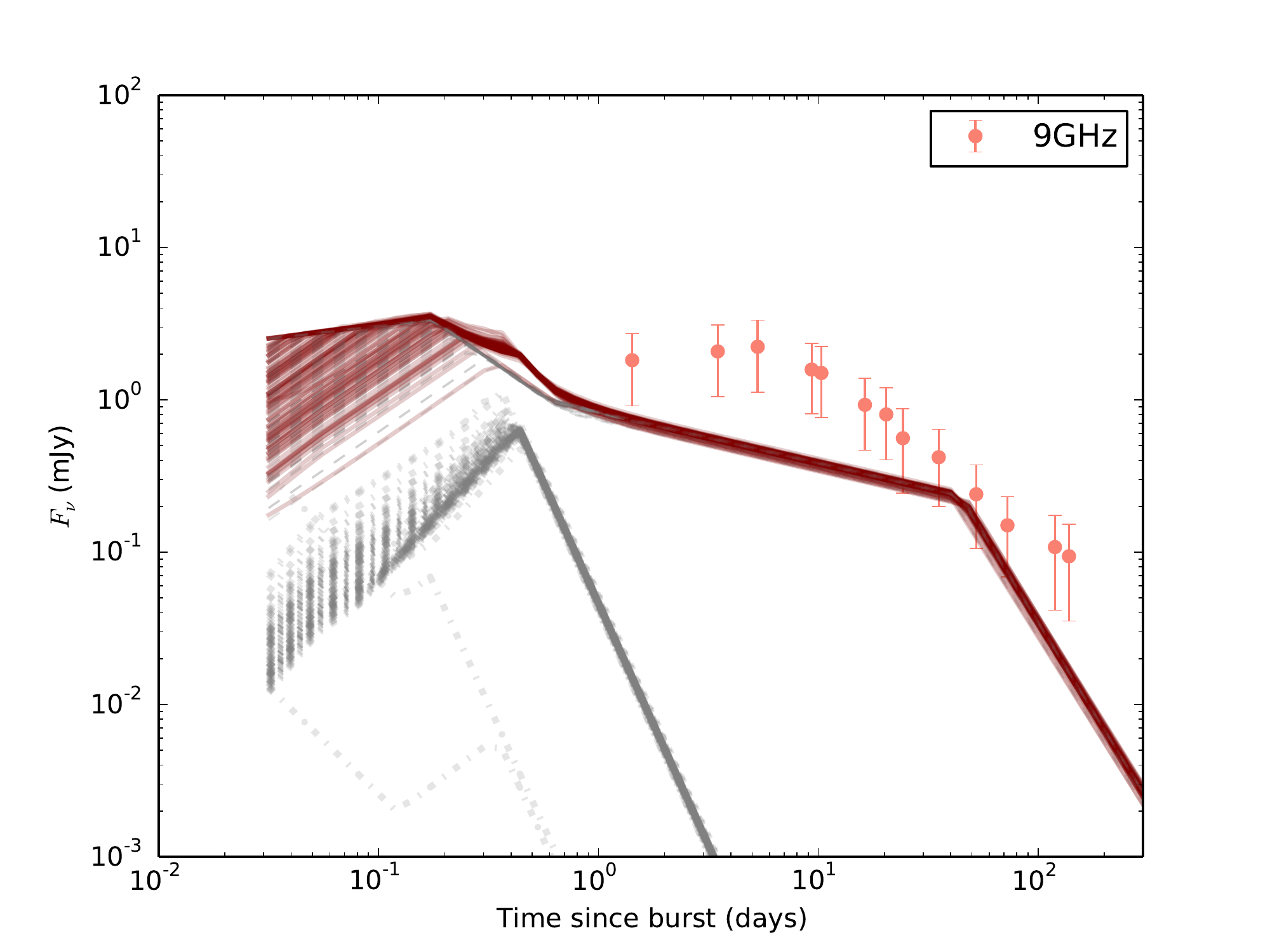}
    \caption{Realisations from our posterior distribution for a constant-density medium. We use the three representative bands to present the results. Gray dashed curves belong to forward shock and gray dotted curves are emission from the reverse shock. Maroon curves represent the total flux. Scintillation is accounted for in the errors of the radio data.} %{\textcolor{red}{May be should add posterior bounds in the caption.}}
    \label{fig:app1}
\end{figure*}
\begin{figure*}
    \centering
    \includegraphics[scale=0.28]{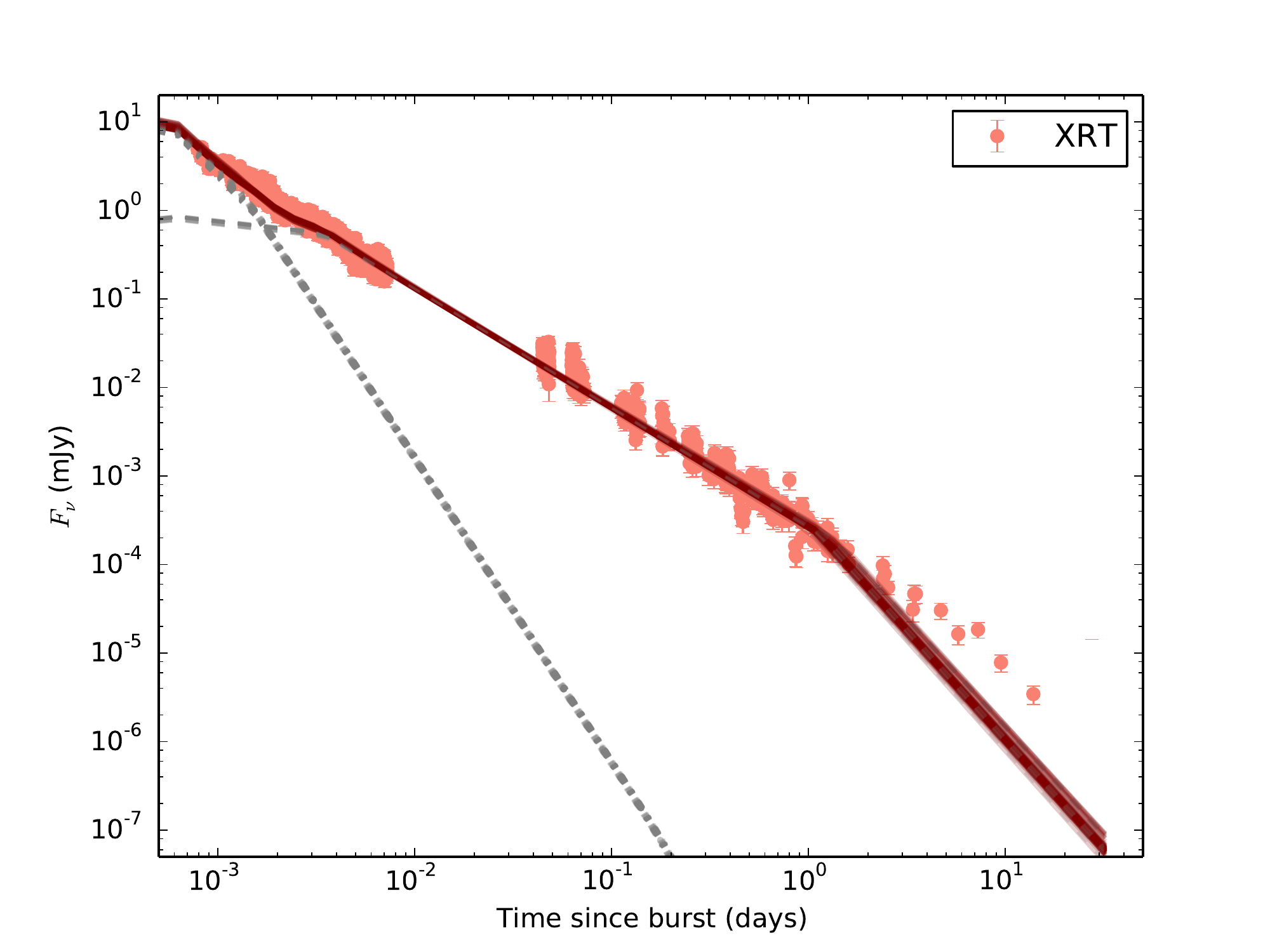}
    \includegraphics[scale=0.28]{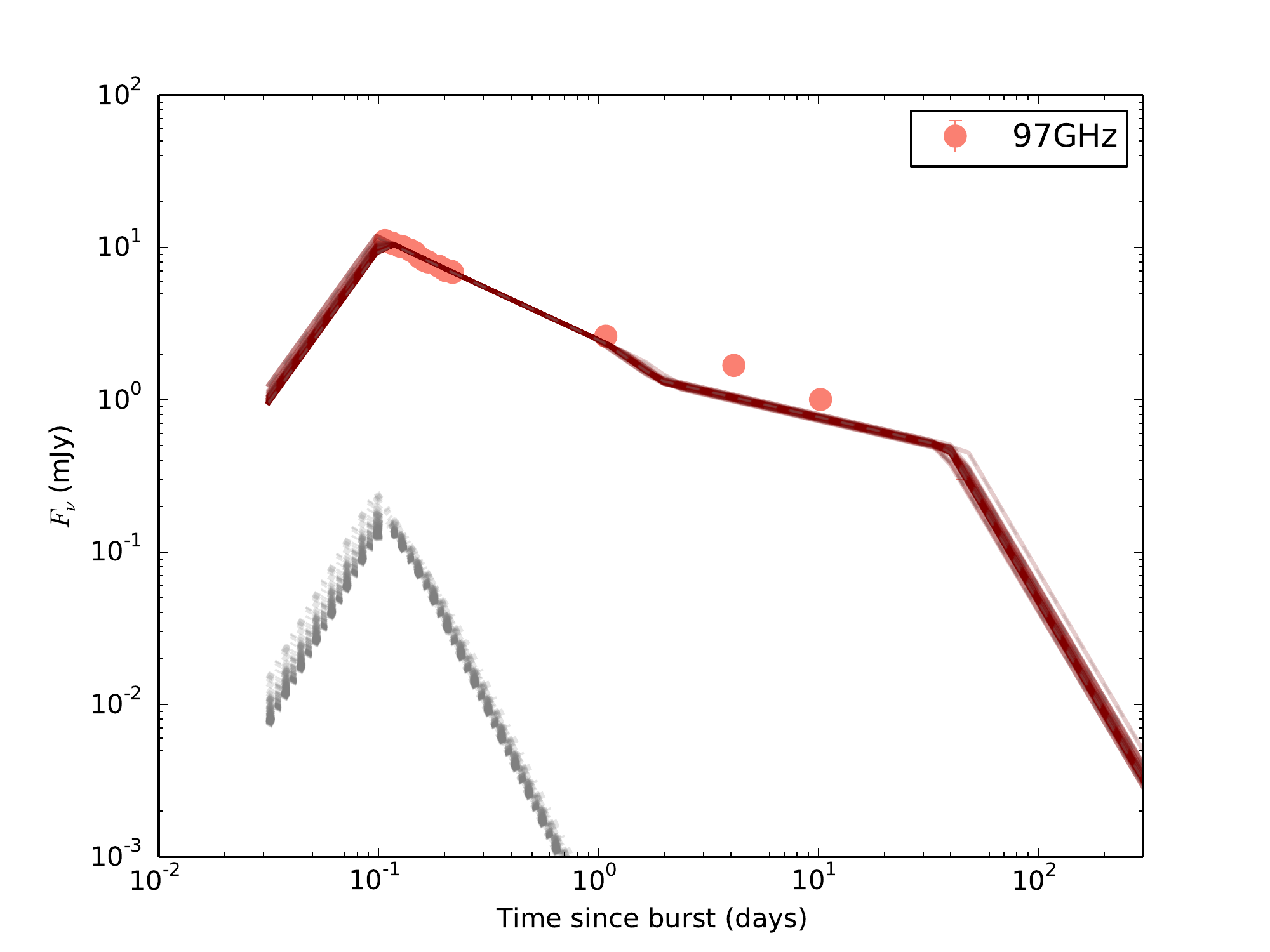}
    \includegraphics[scale=0.28]{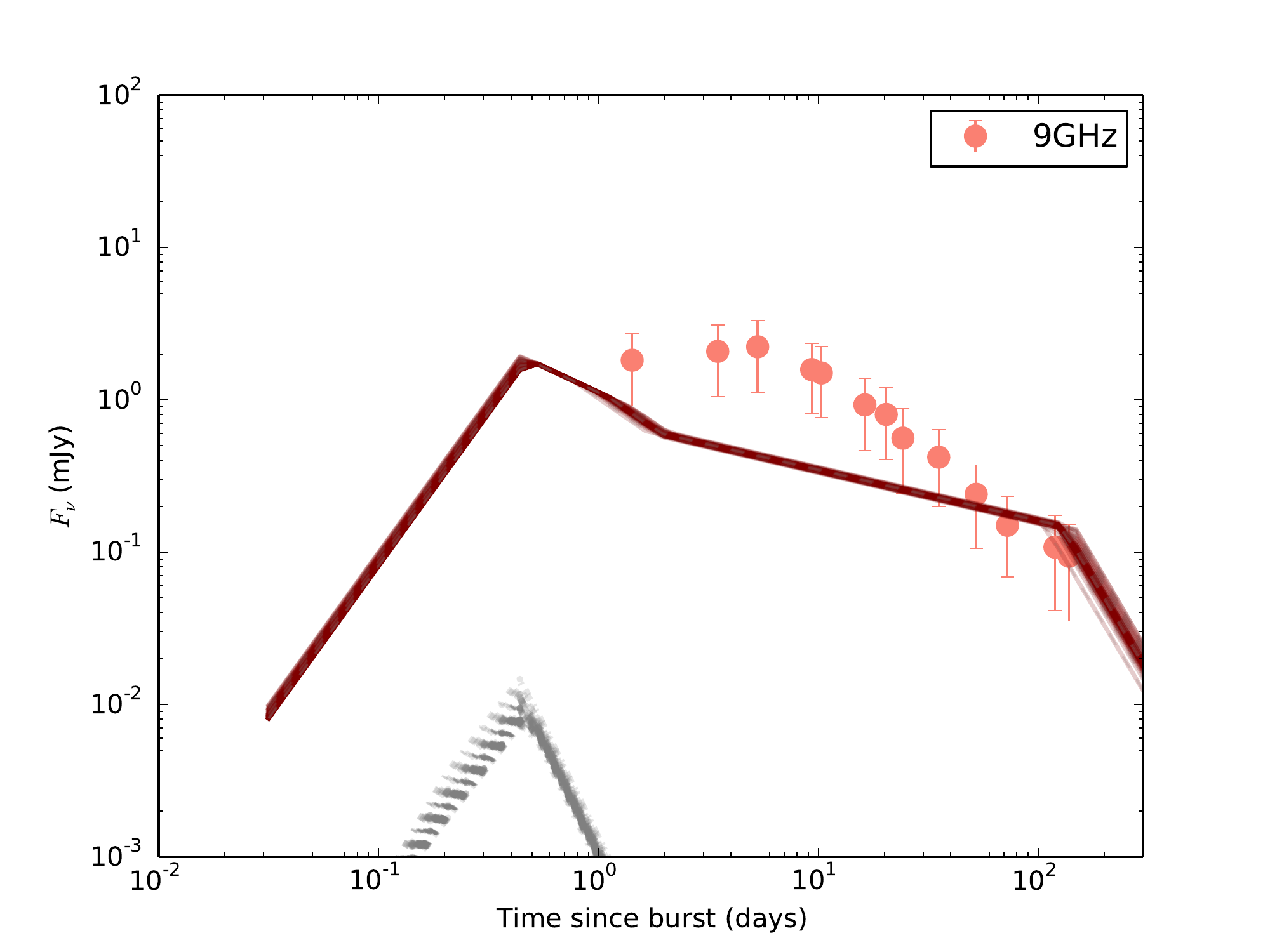}
    \caption{Same as Fig. \ref{fig:app1} but for the wind-blown medium.}
    \label{fig:app2}
\end{figure*}
%
%Reverse shock emission depends, in addition to $E_{\rm iso}$, $\theta_j$, and ambient density, on the initial bulk Lorentz factor $\eta$ of the fireball, electron energy spectrum (characterized by $p_{\rm RS}$) and the fractional energy content in electrons and magnetic field, parametrized as $\epsilon_e^{\rm RS} = \cal{R}_e \epsilon_e$ and $\epsilon_B^{\rm RS} = \cal{R}_B \epsilon_B$ respectively. The post shock-crossing ($t>t_X$) dynamics of the reverse shock is broadly self-similar \citep{Kobayashi:2000af}. However, a range of power-law indices are permissible for the temporal evolution of bulk Lorentz factor $\Gamma_{3}$ of the reverse shocked medium (often referred as medium-3). Therefore, another additional parameter $g$ characterises the evolution of the bulk Lorentz factor of the RS downstream.
%%%%%%%%%%%%%%%%%%%%%%%%%%%%%%%%%%%%%%%%%%%%%%%%%%

% Don't change these lines
\bsp	% typesetting comment
\label{lastpage}
\end{document}